\documentclass[conference]{IEEEtran}
\usepackage{tikz}
\usepackage{amsmath}
\usepackage{amssymb}
\usepackage{tcolorbox}
\usepackage{xspace}
\usepackage{xcolor}
\usepackage{cite}
\tcbuselibrary{most}
\newcommand{\name}{{PingPong}\xspace}
\newcommand{\nameN}{\textsc{Ping}\xspace}
\newcommand{\nameM}{\textsc{Pong}\xspace}

\usepackage{pifont}

\usepackage{subfigure}
\usepackage{booktabs}
\usepackage{threeparttable}
\usepackage{multicol}
\usepackage{multirow}

\usepackage{algorithmicx}
\usepackage{algpseudocode}
\usepackage{enumitem}
\usepackage{listings}
\usepackage{listings-golang} 
\usepackage{colortbl} 
\usepackage{tabularx} 
\usepackage{makecell} 
\usepackage{rotating} 

\usetikzlibrary{
  pgfplots.groupplots,
  matrix,
  patterns,
  chains
}

\usepackage{
  tikz,
  pgfplots,
  pgfplotstable
}

\pgfplotsset{
width=0.465\linewidth,
height=0.4\linewidth,
compat=1.9
} 
\usepgfplotslibrary{groupplots}

\newtheorem{myDef}{Definition}[section]
\newtheorem{myTheorem}{Theorem}[section]
\tcbuselibrary{skins}

\lstdefinestyle{PythonStyle}{
    language=Python,
    basicstyle=\ttfamily\small,
    keywordstyle=\color{green!60!black}\bfseries,
    commentstyle=\color{gray},
    stringstyle=\color{green!60!black},
    showstringspaces=false,
    breaklines=true,
    frame=single,
    backgroundcolor=\color{white},
    morekeywords={Full_Build, Build, Lookup}
}

\definecolor{checkmarkGreen}{rgb}{0.564, 0.933, 0.564} 
\definecolor{crossRed}{rgb}{0.980, 0.502, 0.447} 

\newcolumntype{Y}{>{\centering\arraybackslash}X} 
\newcolumntype{R}{>{\raggedleft\arraybackslash}X} 
\newcolumntype{M}[1]{>{\centering\arraybackslash}m{#1}}

\definecolor{checkmarkGreen}{HTML}{e2fee4} 
\definecolor{crossRed}{HTML}{ffabaa} 

\definecolor{codegreen}{HTML}{0C6427}
\definecolor{codegray}{rgb}{0.5,0.5,0.5}
\definecolor{codepurple}{rgb}{0.58,0,0.82}
\definecolor{backcolour}{rgb}{0.95,0.95,0.92}
\definecolor{codeblue}{rgb}{0.25,0.5,0.5}
\definecolor{commentblue}{HTML}{5696A5}
\definecolor{codered}{HTML}{83210f}
\usepackage{url}
\makeatletter
\def\UrlAlphabet{%
      \do\a\do\b\do\c\do\d\do\e\do\f\do\g\do\h\do\i\do\j%
      \do\k\do\l\do\m\do\n\do\o\do\p\do\q\do\r\do\s\do\t%
      \do\u\do\v\do\w\do\x\do\y\do\z\do\A\do\B\do\C\do\D%
      \do\E\do\F\do\G\do\H\do\I\do\J\do\K\do\L\do\M\do\N%
      \do\O\do\P\do\Q\do\R\do\S\do\T\do\U\do\V\do\W\do\X%
      \do\Y\do\Z}
\def\UrlDigits{\do\1\do\2\do\3\do\4\do\5\do\6\do\7\do\8\do\9\do\0}
\g@addto@macro{\UrlBreaks}{\UrlOrds}
\g@addto@macro{\UrlBreaks}{\UrlAlphabet}
\g@addto@macro{\UrlBreaks}{\UrlDigits}
\makeatother
        
\algnewcommand\NumberedState[1]{%
  \State\stepcounter{ALG@line}\textbf{\arabic{ALG@line}.}~#1%
}

\usepackage{subcaption}
\begin{document}
\title{Metadata-private Messaging without Coordination} 
\author{
 \IEEEauthorblockN{ Peipei Jiang$^{1,2}$, Yihao Wu$^1$, Lei Xu$^3$, Wentao Dong$^2$, Peiyuan Chen$^1$, \\ Yulong Ming$^2$, Cong Wang$^2$, Xiaohua Jia$^2$ and Qian Wang$^{1,}$\textsuperscript{\ding{41}}}
\IEEEauthorblockA{
    $^1$\textit{Wuhan University}, 
    $^2$\textit{City University of Hong Kong}, 
    $^3$\textit{Nanjing University of Science and Technology}}
\IEEEauthorblockA{
    \{pp.jiang, wentao.dong, myl.7\}@my.cityu.edu.hk, \{congwang,csjia\}@cityu.edu.hk\\
    \{yihao.wu, chenpeiyuan, qianwang\}@whu.edu.cn, xuleicrypto@gmail.com
}
}


\maketitle

\setcounter{page}{1}

\begin{abstract}

For those seeking end-to-end private communication free from pervasive metadata tracking and censorship, the Tor network has been the de-facto choice in practice, despite its susceptibility to traffic analysis attacks. Recently, numerous metadata-private messaging proposals have emerged with the aim to surpass Tor in the messaging context by obscuring the relationships between any two messaging buddies, even against global and active attackers. However, most of these systems face an undesirable usability constraint: they require a metadata-private ``dialing'' phase to establish mutual agreement and timing or round coordination before initiating any regular chats among users. This phase is not only resource-intensive but also inflexible, limiting users' ability to manage multiple concurrent conversations seamlessly. For stringent privacy requirement, the often-enforced traffic uniformity further exacerbated the limitations of this roadblock. 

In this paper, we introduce \name{}, a new end-to-end system for metadata-private messaging designed to overcome these limitations. Under the same traffic uniformity requirement,  
\name{} replaces the rigid ``dial-before-converse'' paradigm with a more flexible ``notify-before-retrieval'' workflow. This workflow incorporates a metadata-private notification subsystem, \nameN{}, and a metadata-private message store, \nameM{}. Both \nameN{} and \nameM{} leverage hardware-assisted secure enclaves for performance and operates through a series of customized oblivious algorithms, while meeting the uniformity requirements for metadata protection. By allowing users to switch between conversations on demand, \name{} achieves a level of usability akin to modern instant messaging systems, while also offering improved performance and bandwidth utilization for goodput. We have  built a prototype of \name{} with 32 8-core servers equipped with enclaves and conducted a case study on a real-world messaging metadata dataset to validate our claims. 

\end{abstract}





\section{Introduction}\label{sec:intro}


End-to-end private communication has emerged as an urgent necessity amidst pervasive metadata tracking and censorship~\cite{kill,MayerMM16}. The Tor network~\cite{DingledineMS04} has long been the de facto solution for users seeking anonymity, providing robust protections against many forms of surveillance. However, despite its widespread adoption, Tor remains vulnerable to traffic analysis attacks from even passive observers~\cite{190964,Gilad19}. 

Recently, a new wave of metadata-private messaging systems has emerged to strengthen privacy guarantees and provide end-to-end functionality~\cite{DBLP:conf/ccs/Corrigan-GibbsF10, DBLP:conf/osdi/WolinskyCFJ12,DBLP:conf/uss/Corrigan-GibbsWF13, DBLP:conf/uss/AlexopoulosKT017, sosp15vuvuzela, sosp17atom, osdi18karaoke, sosp19yodel, uss17loopix, DBLP:conf/sigcomm/BlondCCDM15, nsdi20XRD, osdi16pung, riposte15sp, angel2018pir, osdi21Addra, osdi22Groove,nsdi23boomerang,DBLP:conf/ccs/ToveyWG24}. These systems carefully manage observable communication patterns and enforce traffic uniformity, ensuring that metadata does not reveal relationships between communicating parties~\cite{DBLP:conf/sp/AngelKR20,cryptoeprint:2023/313}. Enforcing traffic uniformity across all users is computationally intensive and non-trivial. To achieve traffic uniformity and lightweight obfuscation, many designs rely on an intermediate ``virtual address'' (also referred to as a dead drop) where users can drop and fetch messages. The methods for accessing these virtual addresses obliviously vary across designs. For example, some systems employ differential privacy to protect access patterns~\cite{sosp15vuvuzela,osdi18karaoke,sosp17Stadium,sosp19yodel}, while others use advanced cryptographic techniques~\cite{osdi16pung,angel2018pir,osdi21Addra,DBLP:conf/ccs/ToveyWG24}, or trusted hardware~\cite{nsdi23boomerang}, etc. State-of-the-art systems have demonstrated end-to-end private messaging with various trade-offs in security, performance, and trust assumptions~\cite{nsdi23boomerang}.

While most existing efforts focus on optimizing how messages are exchanged through these virtual addresses (aka the conversation protocol~\cite{DBLP:conf/sp/AngelKR20}), they often assume that 1) the ``location'' of virtual address and 2) the timing of access are determined out-of-band via a separate private dialing system like ~\cite{osdi16alpenhorn}. In other words, users must establish mutual agreement before every conversation and carefully coordinate message exchanges to prevent metadata leakage. However, this ``dial-before-converse'' framework, though effective in securing metadata, is \textit{not flexible} for modern instant messaging experiences, where users expect seamless and asynchronous communication.

In this paper, we revisit this major usability limitation in existing bidirectional metadata-private messaging systems\footnote{Our focus is on bidirectional messaging, which have distinct security requirements compared to unidirectional message delivery~\cite{uss21express} or  broadcast~\cite{nsdi22Spectrum_broadcast} (\S\ref{sec:relatedwork}).}~\cite{sosp15vuvuzela,nsdi20XRD,osdi18karaoke,sosp17Stadium,osdi16alpenhorn,osdi16pung,osdi21Addra,angel2018pir,DBLP:conf/uss/AlexopoulosKT017,nsdi23boomerang}. 
This limitation, largely overlooked until recently~\cite{osdi22Groove,cryptoeprint:2023/313}, arises from the requirement for messaging buddies to coordinate their conversations through a dialing phase. 
To better understand this constraint, we start by briefly reviewing the ``dial-before-converse'' framework, as introduced below.

\begin{figure*}[!t]
    \centering
    \includegraphics[width=0.85\linewidth]{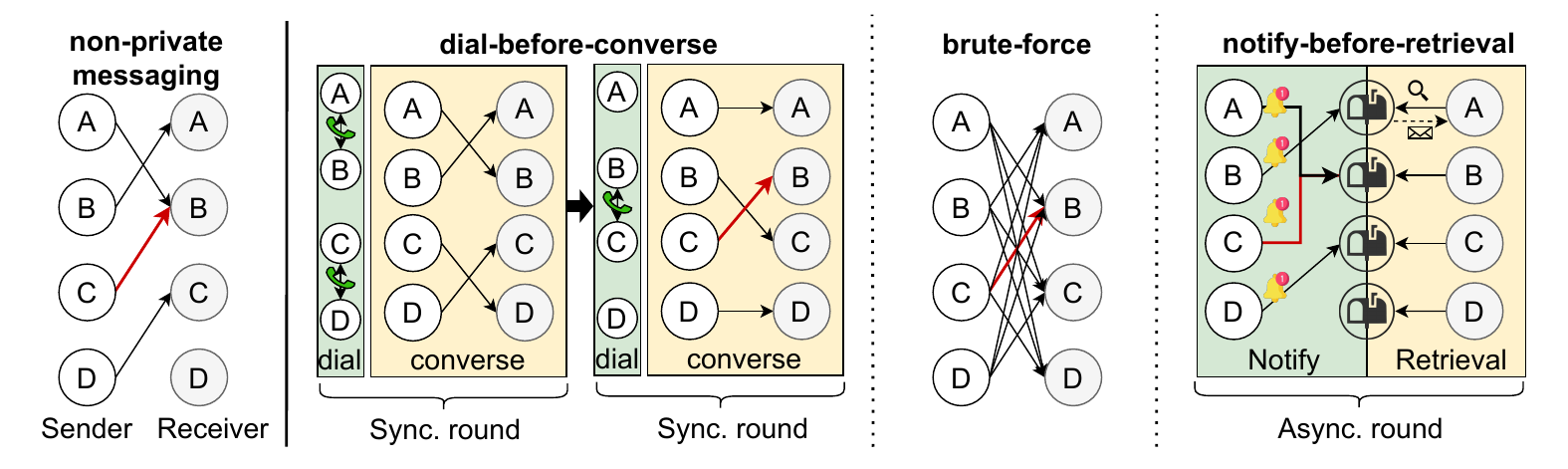}
    \vspace{-2mm}
    \caption{Comparison between the traditional dial-before-converse framework, Groove's brute-force approach and \name{}'s notify-before-retrieval framework.}
    \label{fig:commmodel}
\end{figure*}

\noindent\textbf{Previous framework.}  The ``dial-before-converse'' framework involves add-friend, dialing, and conversation protocol designs~\cite{DBLP:conf/sp/AngelKR20,cryptoeprint:2023/313}. 1) Two users become friends when they establish a shared secret~\cite{DBLP:conf/sp/AngelKR20}, which can be accomplished either out-of-band (e.g., in-person QR code exchange~\cite{uss21express,DBLP:journals/popets/BorisovDG15}) or through an established private contact discovery protocol~\cite{private_contact_discovery}. 2) Any pair of friends intending to converse will coordinate through a costly metadata-private ``dialing'' phase (e.g., one user ``calls'' another) to initiate a session~\cite{sosp15vuvuzela,osdi16alpenhorn,DBLP:conf/sp/AngelKR20,cryptoeprint:2023/313}, where both agree on the same round to join the conversation along with a session key to encrypt the messages. 3) After coordination, the pair of friends will use the conversation protocol to exchange messages in an obfuscated manner, which unlinks sender and receiver but maintains traffic uniformity over all connected clients~\cite{Gilad19,cryptoeprint:2023/313,nsdi23boomerang}. 
Prior efforts have made different design choices to instantiate the common ``dial-before-converse'' framework. However, we argue that this framework  is not the only viable approach for metadata-private messaging and it has two major limitations:

1) \emph{Usability}: A pair of friends is restricted to one conversation at a time~\cite{cryptoeprint:2023/313}. As in Figure~\ref{fig:commmodel}, if Alice is conversing with Bob, she must wait until the end of that conversation before she can  message another friend, Charlie. Similarly, none of Alice's friends can send any messages to Alice, until she finishes her conversation with Bob. In other words, existing conversation protocol under the ``dial-before-converse'' framework does not allow users to transition seamlessly between concurrent conversations, like what they would experience in modern messaging systems.

2) \emph{Cost}: The ``dialing'' protocol adopted in current systems is resource-intensive \cite{osdi22Groove}. For example, Alpenhorn~\cite{osdi16alpenhorn}, a widely used dialing protocol, requires several minutes of coordination between friends for each conversation and consumes tens of gigabytes of user bandwidth per month. This high bandwidth cost arises from its metadata-private design, where users must download a large chunk of mixed ``requests'' and identify the correct one through a trial-and-check process.

As pointed out recently~\cite{DBLP:conf/ccs/ToveyWG24}, besides~\cite{osdi16alpenhorn}, numerous proposals can be adopted to instantiate the dialing protocol~\cite{PS22sec,DBLP:conf/ccs/BeckLM021,OMR22cropto}. But we note that the above constraints still hold, because they arise from the ``dial-before-converse'' framework as a whole. 
The often-enforced traffic uniformity further exacerbates these limitations.

\noindent\textbf{Previous effort in removing dialing.} To our best knowledge, before our work, Groove gave the first and only attempt to address the above limitations~\cite{osdi22Groove}. In order to \textit{remove the dialing phase} before any pairwise conversation, they adopt a brute-force approach that requires every user to exhaustively establish message circuits (or channels) to all possible friends and constantly exchange messages (including cover traffic) on all the circuits\footnote{A similar idea on supporting multiple concurrent conversations has been noted in an earlier private messaging system, Vuvuzela~\cite{sosp15vuvuzela}.}. Since every user is talking to all of her friends concurrently all the time, essentially no dialing between any pairwise users would be needed. While Groove introduces oblivious delegation to mitigate user's cost, this peak-padding method remains expensive in bandwidth and computation cost, imposing severe constraints on friend count, effective throughput (aka goodput), and system scaling. With a max friend count as only 50, Groove needs to maintain 50 million circuits concurrently if there are 1 million clients. 

These observations lead to an immediate question: \emph{is there any alternative viable framework, other than the ``dial-before-converse'' or the above ``brute-force'' approach, that has the same flexibility as instant messaging?}
Ideally, this new framework should function efficiently under the strong traffic uniformity requirement, support much higher goodput, and be capable of horizontal scaling. We answer the question positively with a new system called~\name{}.

\noindent\textbf{Our design intuitions and challenges.}~\name{} draws insights from the purpose of ``dialing'' in previous private messaging systems and the practice of pushing \textit{notifications} in non-private messaging systems. As mentioned earlier, ``dialing'' achieves two goals: 1) \emph{notification}: making a pair of friends notified with each other's intention of conversation, and 2) \emph{coordination}: allowing them to agree on the subsequent conversation rounds details. Among the two goals, we observe that \emph{notification} is more essential, because without this information, a user has no way to know who wants to talk to her and has to go through all of her contacts exhaustively to find the right messages. Meanwhile, the effect of \emph{coordination} essentially binds the pair of friends together, and this ``one-to-one'' binding relationship throughout the conversation protocol appears to be the main cause for the usability limitations mentioned earlier. 

Therefore, in~\name{}, we choose to keep the \emph{notification} while removing the \emph{coordination} that binds any pair of friends. One immediate consequence of this design choice is that messaging exchanges between Alice and her friends are no longer limited to ``one-to-one'' binding, and thus we have to consider more complex many-to-one cases, where many of Alice's friends want to talk to Alice across different conversation rounds. Previous private conversation protocols (e.g., \cite{sosp15vuvuzela,nsdi20XRD,nsdi23boomerang}) under restricted pairwise settings would no longer be applicable here.

\noindent\textbf{Summary of~\name{}.} Building on these insights, our new instant private messaging framework \name{} integrates a metadata-private notification subsystem, \nameN{} (\S\ref{sec:ping}), and a metadata-private message store, \nameM{} (\S\ref{sec:pong}). 
Similar to previous work~\cite{sgxtornsdi,duan2017,SGXboxhan,safebricks,DBLP:conf/ndss/ChenP20,DautermanFDCP21,nsdi23boomerang}, \name{} adopts hardware-assisted secure enclaves~\cite{SinghCQT21,RussinovichCFCD21} for performance, and operates through a series of customized oblivious algorithms and protocols. 
Both \nameN{} and \nameM{} follow the uniformity requirement, and respectively support message notifications and message reads / writes on the message store in an unobservable manner among a large user base. We have formally analyzed the security guarantee following the principle of communication unobservability (See \S\ref{sec:security} and Appendix~\ref{app:proof}).

\name{} achieves much more favorable performance and bandwidth utilization for goodput compared to prior efforts. With 32 8-core servers equipped with secure enclaves, 
\nameN{} achieves sub-second (0.934s) 99th-latency for 50K concurrent clients, and \nameM{}'s throughput reaches 20K fetches/s over a total storage of $2^{29}$ messages (\S\ref{sec:evaluation}). We also show why our new framework is more efficient than the traditional dial-before-conversation framework via a case study on the real-world messaging metadata dataset, CollegeMsg (Appendix \ref{sec:evaluation_framework}). 


 \section{Threat Model and Goals}\label{sec:overview}

\name{} is a metadata-private instant messaging system that supports flexible end-to-end messaging~\cite{osdi22Groove} and communication unobservability~\cite{cryptoeprint:2023/313,KuhnBSJS19}. 
\name{} relies on hardware trust, and provides metadata privacy with cryptographic guarantees~\cite{sosp17atom,nsdi20XRD,nsdi23boomerang} (in contrast to those rely on differential privacy~\cite{osdi22Groove,osdi18karaoke,sosp17Stadium,sosp15vuvuzela}).
Therefore, we consider two key threat models: 1) the secure enclave threat model and 2) the metadata-private messaging threat model.

\subsection{Enclave Threat Model and Guarantees}

\name{} is designed upon general secure enclaves, including the literature advanced ones~\cite{BourgeatLWZAD19,CostanLD16,SinghCQT21,RussinovichCFCD21}, and in our implementation we build \name{} with Intel SGX. 
Abstractly, we assume an uncompromised enclave provides the following security properties:


\textit{1) Integrity:} The enclaves support application-level attestation, allowing the validation of codes and applications through standard remote attestation~\cite{RA-TLS,ratls_code,ra_api} and multi-enclave attestations~\cite{VC3_SP15,safebricks,duan2017,Apache_Teaclave,chen2022mage}. Therefore, the code is assumed to be correct and faithfully follow the algorithms. Additionally, we assume that keys and secure connections can be established using standard practices, e.g., public key infrastructure and remote attestation.
    
\textit{2) Confidentiality:} The memory data running inside the secure enclaves is encrypted and hardware-enforced to prevent access by non-enclave applications. However, the memory access pattern is not protected, which can be leveraged to extract secrets or recover the private data, e.g., through side channels like cache timing~\cite{BrasserMDKCS17,HahnelCP17,MoghimiIE17}, paging~\cite{BulckWKPS17}, and memory bus snooping~\cite{DBLP:conf/uss/LeeJFTP20}, etc. 

As we assume a typical threat model for secure enclaves that leak the memory access pattern~\cite{MishraPCCP18, DautermanFDCP21, DBLP:conf/isca/McKeenABRSSS13, sgxtornsdi,nsdi23boomerang}, \name{} incorporates general-purpose oblivious primitives for both external and internal memory and carefully manages control flow branches, ensuring that the memory access pattern is data-independent and control-flow oblivious~\cite{SasyJG22}. 

\noindent\textbf{Attacks out of scope.} \name{} does not implement specific defenses against rollback attacks~\cite{DBLP:conf/sp/ParnoLDMM11} and other attacks based on power consumption channel~\cite{DBLP:conf/sp/MurdockOGBGP20,DBLP:conf/uss/ChenVMDOG21} and transient execution~\cite{DBLP:conf/sp/RagabMRBG21,DBLP:conf/sp/BulckM0LMGYSGP20}. For software rollback threats, memory sealing can be avoided by keeping all data within the enclave, as suggested in~\cite{DBLP:conf/osdi/ConnellFSDP24}. However, this mitigation will limit the available storage space since all data should remain in the enclaves. Additionally, we assume that the compilation process will not interfere with the designed control-flow obliviousness.



\subsection{Metadata-private Messaging Threat Model}
Previous work on metadata-private messaging systems typically assumes both passive and active global attackers~\cite{Gilad19,sosp19yodel,nsdi23boomerang,osdi18karaoke,osdi22Groove}, who can observe or interfere with global network traffic to infer communication relationships and status. 
\name{} follows this threat model. 
Specifically, passive attackers can observe the group of clients connected to the system, as well as the volume and timing of the packets sent and received by each client. The active attackers can adaptively block, inject, or replace traffic from and to the clients. Further, we assume the attackers can control up to $m-2$ clients out of a total of $m$ clients.  
A malicious/compromised friend could craft junk notifications/messages, causing a denial-of-service attack on the target client. 
Our system ensures that, in such cases, the privacy of the remaining uncompromised clients is preserved.

The active interactions with a compromised friend are inevitably exposed to the attacker controlling the friend's account.
We consider leakage through compromised friends~\cite{DBLP:conf/sp/AngelKR20,wpes18Angel} an orthogonal and complementary privacy issue, where a compromised friend may infer the status of a victim from his/her response behaviors. As a direct mitigation, one could integrate a private resource allocator~\cite{DBLP:conf/sp/AngelKR20} at the client side on top of existing metadata-private messaging designs to address this privacy leakage.

%

\subsection{Goals and Security Notions} \label{sec:commmodel}
\noindent\textbf{Usability goal: Instant messaging without coordination.}
Our goal is to make metadata-private messaging align more closely with the intuitive messaging habits. 
Specifically, \name{} allows users to chat without: 1) needing to coordinate with their contacts for every conversation, and 2) being enforced to stay in lasting conversation sessions. 
\name{} provides message storage for asynchronous online status, enabling buddies to send messages and retrieve them later without revealing the correlationship.
While \name{} does not require a dialing phase for messaging, it does require an initial ``add-friend'' phase, where users become friends and exchange necessary secrets at the first place. Note that this phase is also a prerequisite in many non-private instant messaging applications that requires access control. 
Fortunately, it is a \textit{one-time process} that can be completed either offline (e.g., via QR code exchange~\cite{uss21express}) or online through a metadata-private add-friend system~\cite{osdi16alpenhorn}.


\noindent\textbf{Privacy goal: Communication unobservability with cryptographic guarantee.} \name{} aims to achieve \textit{communication unobservability}, a well-known concept in anonymous communication~\cite{HeviaM08,terminologyunoberser,KuhnBSJS19}, as done by prior  systems~\cite{sosp15vuvuzela,osdi16pung,sosp17Stadium,osdi22Groove}, but with \textit{cryptographic privacy}~\cite{nsdi20XRD,osdi21Addra,osdi16pung,nsdi23boomerang}. 
Specifically, \name{} ensures that whether any pair of clients in the system is engaged in a conversation remains \textit{indistinguishable} to potential attackers.
As conversations in \name{} are uncoordinated, the communication and privacy aspects require additional modeling. 
We formally demonstrate this privacy guarantee by modeling the communication traces, which captures the nuances of how messages are exchanged asynchronously across multiple rounds. 

\noindent\textbf{Security notions.}
Modeling security notions for uncoordinated messaging needs more twists compared to prior pairwise systems. 
Within an instant messaging system, the interactions among clients can be conceptualized as a series of communication events, where messages are passed from one party to another. We refer to such instances of message transmission as a communication trace.

\begin{myDef}[Communication Trace] 
Let $C = \{C_1, \dots, C_N\}$ represent a set of clients connected to the messaging system, and $\mathcal{M} = \{m_1, \dots, m_{\ell}\}$ denote a set of messages intended for delivery. The communication trace is defined as a three-tuple $(C_i, m_k, C_j)$, where the sender $C_i$ transmits a message $m_k$ to the message store, and the receiver $C_j$ retrieves it subsequently.
\end{myDef}


With the communication trace, we now introduce the notion of indistinguishability for metadata-private instant messaging.

\begin{myDef}[Communication Trace Indistinguishability]\label{def:ctu}(Informal) Let $\mathsf{IM}$ be a  instant messaging system with security parameter $\lambda$, and let $T = \{t_1, \dots, t_{\ell}\}$ represent the communication trace established among a set of clients $C = \{C_1, \dots, C_N\}$ and a set of messages $\mathcal{M} = \{m_1, \dots, m_{\ell}\}$. 
We say that an instant messaging system $\mathsf{IM}$ achieves \textit{communication trace indistinguishability} if, for any message $m_i \in \mathcal{M}$ and any two pairs of clients $(C_i, C_j)$ and $(C_i^{\prime}, C_j^{\prime})$, the probability that the communication trace $t_i$ associated with $m_i$ occurs satisfies  
\begin{equation*}
|{\Pr}[t_i = (C_i, m_i, C_j)] - {\Pr}[t_i = (C_i^{\prime}, m_i, C_j^{\prime})]| < \mathsf{negl}(\lambda),
\end{equation*}
in the view of a probabilistic polynomial-time attacker, where $\mathsf{negl}(\lambda)$ is a negligible function dependent on $\lambda$.
\end{myDef}

This notion ensures that, from an attacker's perspective, all potential sender-receiver pairs related to a given message are equally likely. As a result, any attempt to distinguish between different communication traces based on observed traffic patterns becomes ineffective.
Notably, this property inherently provides sender and receiver anonymity, preventing the attacker from determining which client sends or receives a message. 

\subsection{Oblivious Primitives}\label{sec:background}\label{sec:primitive}

To hide memory access pattern within enclaves, \name{} adopts the following key oblivious building blocks.

\noindent\textbf{Oblivious assignment} can conditionally assign elements without leaking the condition. This can be implemented in x86-64 assembly, which operates solely on registers~\cite{DBLP:conf/uss/OhrimenkoSFMNVC16}. 
With these fundamental primitives, our system utilizes high-level functions such as Oblivious Choose and Oblivious Equal, as available in the oblivious primitive library~\cite{sgboostres,DBLP:conf/ccs/LawLPPSSYZZ20}. We employ $\texttt{OblChoose}(\text{flag}, a, b)$, which returns the value of $a$ if the flag is true, and $b$ otherwise, and $\texttt{OblEqual}(a, b)$, which returns true if $a$ equals $b$.

\noindent\textbf{Oblivious sort} ($\texttt{OSort}$) can sort data according to their keys while reveals nothing about their order in the original sequence~\cite{DBLP:conf/soda/AsharovCNP0S20,SasyJG22,Sasy0G23}. Our implementation uses bitonic sort~\cite{Batcher68}, which sorts in a predefined order through compare-and-swap operations. Although it has a complexity of $O(n\log^2n)$, where $n$ is the length of the sequence, its parallelizability enhances its efficiency. We present this function as $\mathsf{seq}.\texttt{OSort}(\mathsf{key})$, meaning the sequence is sorted based on the keys.

\noindent\textbf{Oblivious compaction} can compact a sequence by preserving items tagged with $\mathsf{true}$ while filtering out those with $\mathsf{false}$ without revealing their original positions. We use Goodrich's compact function~\cite{DBLP:conf/spaa/Goodrich11} with $O(n\log n)$ complexity, and denote it as $\mathsf{seq}.\texttt{OCompact}(\mathsf{tag})$.

\section{\name{} overview}\label{sec:pingpong} \label{sec:overview}

\subsection{The Notify-before-retrieval Framework}\label{sec:mainworkflow}

\name{} is a metadata-private messaging system that avoids the overhead of costly dialing protocols and rigid synchronous exchanges. 
Unlike prior systems that rely on costly coordination mechanisms for metadata privacy, \name{} provides a lightweight and user-friendly solution by tackling two key challenges: 1) managing conversations without any coordination in advance, and 2) providing a message buffer for asynchronous message delivery. 

To address these challenges, \name{} introduces a new notify-before-retrieval framework, which decouples notifications from message retrieval. This framework relies on two core components:

\begin{itemize}
    \item \nameN{} (\S\ref{sec:ping}): A metadata-private notification subsystem that ensures \textit{traffic uniformity} by delivering a consistent volume of data to all users, regardless of actual message activity, in a lightweight and non-interactive manner.
    \item \nameM{} (\S\ref{sec:pong}): An oblivious message store that guarantees \textit{unlinkability} between asynchronous message writes and reads, supporting flexible delivery across different rounds.
\end{itemize}

\begin{figure*}[t]
    \centering
    \includegraphics[width=0.95\linewidth]{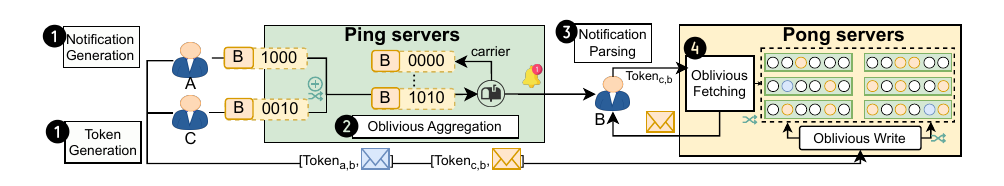}
    \vspace{-2mm}
    \caption{Overview of \name{}. Users A and C simultaneously send messages to user B. }
    \label{fig:overview}
\end{figure*}

Figure~\ref{fig:overview} shows the architecture and components of \name{}: a set of clients, \nameN{} system, and  \nameM{} system. 
Both subsystems are horizontally scalable among distributed servers equipped with secure enclaves, while clients operate without the need for enclave support.
Upon establishing an online connection, clients engage in two  loops of interactions with the \nameN{} and \nameM{} servers as described below
\footnote{For space considerations, the pseudocode for the client main loop is deferred to Figure~\ref{fig:clientloop} in Appendix~\ref{sec:cltloop}.}.


\noindent\textbf{Sending a message.}
The user begins by popping up a message from the message queue and generating a notification packet and a message packet, by invoking the \texttt{GenNotf} function for notification packets and the \texttt{GenMsg} function for message packets (\S\ref{sec:generatingnotf}). Once prepared, these packets are sent to the respective \nameN{} and \nameM{} servers (\ding{202}).
Upon receiving a notification batch, \nameN servers obliviously aggregate the notifications and generate a notification digest for each recipient (\ding{203}). 
Meanwhile, \nameM servers write the message packets to the oblivious  store. 


\noindent\textbf{Fetching a message.}
Upon receiving the notification digest, the user decodes the digest to retrieval tokens (\ding{204}), with \texttt{ParseNotf} function (\S\ref{sec:generatingnotf}). These tokens are then queued for scheduling (\S\ref{sec:schedule}).
The user then pops up a token from this queue and generates a retrieval request to \nameM{} (\ding{205}). 

\subsection{Technical Overview for \nameN{} and \nameM{}}
\noindent\textbf{\nameN{}: Metadata-private notification} (\S\ref{sec:ping}). 
In initiating conversations, we have to consider the ``many-to-one'' scenario, where many of Alice's friends may send notifications to Alice, especially within the same round. 
Instead of maximum padding~\cite{osdi22Groove}, which imposes unnecessary computational and communication overhead, \nameN{} compacts the notifications. By encoding notifications using one-hot vectors, we compact the many-to-one scenario into a one-to-one representation. We design an oblivious aggregation framework that compresses all notifications for a recipient into a single bit-vector. This approach eliminates the need for costly padding while preserving metadata privacy with lightweight computation.

It is worth noting that, prior solutions often rely heavily on dialing protocols to coordinate conversations~\cite{sosp15vuvuzela,osdi16pung,osdi21Addra,DBLP:conf/ccs/ToveyWG24,DBLP:conf/sp/AngelKR20}. These protocols require explicit interactions, such as sending invitations and waiting for acknowledgments~\cite[\S6]{DBLP:conf/ccs/ToveyWG24}. In contrast, notifications in \nameN{} are designed to allow individual users to independently organize their retrievals, without the need for such coordination.

\noindent\textbf{\nameM{}: Metadata-private message store} (\S\ref{sec:pong}).
With notifications in place, the focus narrows down to achieving per-message unlinkability within the message store, i.e., unlink the writes and reads of the messages. Unlike most metadata-private messaging systems that typically support only in-round message exchanges~\cite{nsdi23boomerang,nsdi20XRD,osdi18karaoke}—a limitation that enforces strict user coordination—\nameM{} introduces a more flexible approach. It enables message delivery (i.e., oblivious writes and retrievals) across different rounds, eliminating the need for immediate message exchanges.
This relaxation mirrors the seamless user experience of modern messaging systems, allowing users to transition fluidly between conversations without rigid time constraints. By decoupling message delivery from strict in-round exchanges, \nameM{} removes the coordination bottleneck often inherent in traditional systems.

While \nameM{} can be instantiated using general ORAM designs implemented on secure enclaves~\cite{DautermanFDCP21,MishraPCCP18,TinocoGS23}, we develop a specialized oblivious store to achieve better runtime performance. This design leverages oblivious hash tables and introduces customized optimizations, such as hierarchical storage and deamortization.

For clarity of presentation, we will introduce our designs in a single-server model in \S\ref{sec:ping} and \S\ref{sec:pong}, and discuss how to further make these systems \textit{horizontally scalable} in \S\ref{sec:HSarchi}.





\subsection{User Schedule}\label{sec:schedule}
\noindent\textbf{Online users}. 
To hide the real communication patterns against traffic analysis attacks, online users should maintain their own fixed interaction pattern with the system, independent of the real conversation status. 
For example, in Figure~\ref{fig:overview}, user B needs to retrieve user C's message in the next round if he opts to retrieve one message per round.
Irregular online patterns may expose users to long-term intersection threats~\cite{DBLP:conf/pet/BertholdL02}.


\noindent\textbf{Offline users}. In the case of offline users, the system can adopt a proxy delegation mechanism as introduced in \cite{osdi22Groove}. Specifically, when a user logs in, he/she initiates a connection to a node within the \nameN{} server cluster. This node is then responsible for collecting notification digest(s) for the user, irrespective of their online status. The user would still need to use a fixed (but more relaxed) rate to interact with the proxy. When the user re-establishes her online presence again, she retrieves the stored notification digests from the proxy during her offline period. 


\section{\nameN{}: Metadata-private notification}\label{sec:ping}

\nameN is a lightweight metadata-private notification system that enables users to signal unread messages to their friends, who can then arrange their own fetching schedule to ensure uniform communication patterns critical for metadata privacy.
Unlike prior systems relying on global broadcasts of dialing invitations~\cite{osdi16alpenhorn}, \nameN{} employs compact one-hot bit-vectors for efficient notification representation.
While bit-vectors are a known method for compact encoding~\cite{toubiana2010adnostic}, their use in our scenario demands careful design. Taking bit-vectors as a starting point, \nameN{} addresses three key challenges: (1) preventing ``spam'' through tamper-proof notifications, (2) enabling non-interactive message coordination, and (3) ensuring volume-consistent notification aggregation to preserve traffic uniformity. 

\subsection{Client: Notification Encoding and Decoding}  \label{sec:pingclt} \label{sec:generatingnotf}

\nameN{} leverages bit vectors for efficient notifications, with two critical design considerations to ensure security and usability: preventing notification flooding and supporting non-interactive message coordination.
Figure~\ref{fig:clientfunccode} illustrates three main functions of local client  operations, detailing how a user can locally prepare the notification packets (as a sender), and parse the vectors to learn where to fetch the corresponding messages (as a receiver). 

\noindent\textbf{Notification preparation: Sealing against flooding.} 
\nameN{} encodes sender identities using one-hot bit-vectors, where each bit corresponds to a friend in the receiver's list. 
However, if senders could freely craft these vectors, a malicious sender might falsely impersonate another user or set all bits to 1, thereby triggering a flooding attack by causing numerous nonexistent message retrievals at very low cost.

\nameN addresses this threat with the design of sealed notification tokens (\texttt{NotfToken}). 
Instead of having the senders prepare the bit vectors, the receiver generates a sealed token. Specifically, the receiver uses the enclave's public key (\texttt{PingPKey}) to encrypt the vector (together with a label for server-side aggregation) and shares it with the sender once they become friends.
This token is stored in the sender's local \texttt{FriendList}\footnote{\texttt{FriendList} is a structured list mapping each friend to their identifier (\texttt{ID}), secret key (\texttt{sk}), and a counter for non-interactive coordination.}. The token generation process is as follows:
\vspace{-1mm}
\begin{lstlisting}[basewidth=0.5em]
Idx := FriendList.AddFriend(ID)
NotfVec := [0] * MAX_FRIENDS
NotfVec[Idx] = 1
NotfToken := Encrypt((NotfVec, mylabel), PingPKey)
\end{lstlisting}
\vspace{-1mm}
This approach has a similar flavor to access control encryption~\cite{DBLP:conf/tcc/DamgardHO16}: it enforces fine-grained control over where a sender can write.  
By assigning a unique token to each friend, the system ensures that senders can only modify their own associated bit, preventing them from tampering with bits belonging to others. Once the token stored, the sender can use this token to generate notification packets efficiently (see \texttt{GenNotf}, Lines 1-6 in Figure~\ref{fig:clientfunccode}).

\noindent\textbf{Notification parsing: Non-interactive message coordination.}
The core challenge in \name{} is enabling senders and receivers to independently determine message locations without real-time coordination. 
\nameN addresses this by designing retrieval tokens that can be derived independently by both the sender and the receiver using a shared secret key (\texttt{sk}) and an updatable counter.
With notification vector indicating the sender identities, the receiver can retrieve the ID, \texttt{sk} and the counter from the local friend list. 
A pseudorandom function (PRF) then processes their IDs and the counter (incremented per message) to produce unique yet predictable tokens, enabling both parties to locate messages seamlessly across rounds.
This process involves two key functions:
\begin{itemize}
    \item \texttt{GenMsg} (Lines 7-17): Generates message packets with retrieval tokens and encrypted content.
    \item \texttt{ParseNotf} (Lines 18-27): Decodes notification vectors to retrieval tokens that can pinpoint message locations.
\end{itemize}

\begin{figure}[t]
\begin{tcolorbox}[
    enhanced,
    colback=white, 
    colframe=white, 
    boxrule=0pt, 
    left=0pt, 
    right=0pt, 
    top=-12pt, 
    bottom=0pt, 
  ]
\begin{lstlisting}[basewidth=0.5em]
def GenNotf(buddy):
  if buddy is not None:
    notfPkt = Encrypt(buddy.NotfToken, PingKey)
  else:  # For idle client, generate a blank vector
    notfPkt = Encrypt(None, PingKey)
  return notfPkt
def GenMsg(buddy, msg):
  if buddy is not None:
    sk = buddy.sk; counter = buddy.counter
    token = PRF(sk, buddy.ID + myID + str(counter))
    val = Encrypt(msg, sk)
  else:
    token = DUMMY; val = random_bytes()
  msgPkt = Encrypt((token, val), PongKey)
  # Note: update the counter only for successful delivery
  if buddy is not None:   buddy.counter += 1
  return msgPkt
def ParseNotf(NotfVec):
  msgTokens = []
  for b, Nbit in enumerate(NotfVec):
    if Nbit == 1:
      buddy = FriendList[b]
      counter = buddy.counter, sk = buddy.sk
      token = PRF(sk, myID + buddy.ID + str(counter))
      FriendList[b].counter += 1
      msgTokens.append(token)
  return msgTokens
\end{lstlisting}

\end{tcolorbox}
\vspace{-5mm}
  \caption{Client local functions. For simplicity, we omit the definitions of standard functions such as Encrypt, PRF, etc. }
    \label{fig:clientfunccode}
\end{figure}

\noindent\textbf{\textit{Remark}: Retrieval schedule.}\label{sec:retrievalSchedule}
After parsing notifications into tokens, users can customize retrieval schedules to their preferences—e.g., prioritizing tokens from key friends or adopting a first-in-first-out approach. The client can also count the unfetched messages for each buddy, which is similar to displaying counts of unread messages for notification in instant messaging applications. 

To further improve communication efficiency, \nameN{} can be extended to use one notification to indicate multiple messages. Once a conversation 
is initiated, future conversations can be indicated during the current conversation. This is useful when sending a burst of messages to a friend without multiple notification packets.

\subsection{Server: Oblivious Notification Aggregation}  \label{sec:pingserver} 

Like many prior metadata-private message exchange systems~\cite{osdi18karaoke,sosp15vuvuzela,sosp17Stadium,osdi22Groove}, \nameN{} operates in rounds. In each round, \nameN{} servers aggregate notification vectors from senders and generate a notification digest for each client. 
Figure~\ref{fig:aggrcode} shows the \nameN{} server operations in one round. 

Below, we first present the high-level insight behind our oblivious aggregation approach, which ensures data-independent memory access patterns, and then detail the step-by-step server operations.  

\noindent\textbf{Insight: Carrier messages for uniform traffic.} 
The primary challenge in \nameN{} is securely aggregating bit vectors and delivering notification digests to clients without leaking metadata.
Unlike the pair-wise model where clients bidirectionally exchange messages, uncoordinated communication exhibits asymmetrical user behavior, such as users being offline (absent receiver) or idle (absent sender). 
For instance, simply compacting incoming vectors could leave idle clients without receiving any digests, revealing their inactivity.  


To address this, we introduce the design of \textit{carrier messages}: server-generated placeholders with all-zero bit vectors and client-specific labels.
These ensure every client receives at least one digest each round, regardless of actual notification pattern, maintaining consistent traffic volumes.
The carrier message has two roles: 1) It helps establish a bidirectional communication structure via labels, simplifying adaptation to scalable, pairwise-like exchanges; 2) It acts as placeholders for idle clients, concealing their lack of notifications.  

\vspace{10pt}

Below are step-by-step operations for \nameN{} servers. 

\noindent\textbf{Step 1: Processing notification packets.}
This step involves decrypting and parsing the received notification packets (Lines 1-8). 
Once the decryption is complete, the enclave proceeds to unseal the notification tokens, by decrypting the tokens using a secret key, which is securely shared among all \nameN enclaves. 
The secure enclave should also perform an essential function of filtering out expired messages from previous round by checking the timestamp of each packet. This is a standard procedure in prior work~\cite{osdi18karaoke,nsdi23boomerang}, and for brevity, it is omitted from the pseudocodes.

\noindent\textbf{Step 2: Preparing carrier messages.}
As discussed, we introduce a carrier message-based notification exchange mechanism to complete the access pattern  (Lines 9-11). 
To ensure data integrity, we let the server generate carrier messages for all clients registered on it.
These carrier messages, comprised of all-zero bit-vectors and sharing the same \texttt{label} as notifications, are designed to ``absorb'' all notification vectors at the virtual address and ``carry'' such information to clients. 

\noindent\textbf{Steps 3\&4: Oblivious aggregation.} \label{sec:aggregation}
This step involves the bitwise OR of bit vectors sharing the same labels into a single vector to be carried by a carrier message. 
This process is executed within the enclaves, and memory access pattern are carefully protected by oblivious sorting and a linear scan.
Initially, packets are obliviously sorted by their labels and carrier tags (Line 14).
Following this, packets tagged with the same label are sequentially aligned, with the carrier message positioned at the end of each group.
The server then scans this sequence, aggregating vectors with the same label onto the most recently accessed vector. Specifically, if the current packet shares a label with the preceding one, it implies they belong to the same group (Line 18). The server then updates the current vector with the aggregation of the previous one (Lines 19-21).  This step employs an oblivious choose function to ensure that the adjacency relationship of vectors, and consequently, the count of each label remains confidential.

\noindent\textbf{Step 5: Generating notification digest.} \label{sec:generatingdigest}
By oblivious aggregation, all notification vectors for each client are expected to be compiled into the carrier message. Note that, the positions of each carrier message within the sequence remain undisclosed during the whole process. 
Therefore,  the next step is to extract the carrier message from the sequence. 
We employ an oblivious compact function~\cite{DBLP:conf/spaa/Goodrich11} on the sequence, using the flag  \texttt{is\_carrier}. Then, we can obtain the carrier message, i.e., the notification digest, for each client, without exposing the position of it. 
Finally, the delivery of the notification digest depends on the client's availability:
If the client is online, the server sends the digest directly to the client by replying the RPC calls. If the client is offline, the server forwards the digest to a delegated proxy (\S\ref{sec:schedule}).

This approach, which involves sending only the carrier message containing aggregated information from multiple notifications, ensures that the only public information disclosed is the total number of clients receiving messages. This preserves individual notification details and the confidentiality of notification relationships.


\begin{figure}[t]
\begin{tcolorbox}[
    enhanced,
    colback=white, 
    colframe=white, 
    boxrule=1pt, 
    left=0pt, 
    right=1pt, 
    top=-10pt, 
    bottom=1pt, 
  ]
\begin{lstlisting}[basewidth=0.5em]
def PingServerRound(recv_notfs):
# Step 1: Decrypt and parse the notifications
  pkts = []
  for (clt, notf) in recv_notfs: 
    NotfToken = Decrpt(notf, SessionPingKey[clt])
    pkt.NotfVec, pkt.label = Decrypt(NotfToken, PingSKey)
    pkt.is_carrier = 0
    pkts.append(pkt) 
# Step 2: Create one carrier message for each client
# whose is_carrier = 1, NotfVec = [0] * MAX_FRIENDS
  carriers = CreateCarriers(ClientSet) 
  pkts.append(carriers)
# Step 3: OSort pkts by label, breaking tie: is_carrier
  pkts.OSort(key=(pkt.label, pkt.is_carrier))
# Step 4: Oblivious notification aggregation
  for pkt in pkts:
    # Check if the current label repeats its prev neighbor
    is_rep = OblEqual(pkt.label, Prev(pkt).label)
    agg_vec = Prev(pkt).NotfVec | pkt.NotfVec # bitwise OR
    # Update NotfVec by agg_vec if it repeats its neighbor
    pkt.NotfVec = OblChoose(is_rep, agg_vec, pkt.NotfVec)
# Step 5: Obliviously extract carrier messages
  pkts.OCompact(key = is_carrier)
# Step 6: Send the notification digests
  SendNotfToClients(sockets, pkts)  
    
\end{lstlisting}

\end{tcolorbox}
\vspace{-5mm}
  \caption{\nameN{} server operations.}
    \label{fig:aggrcode}
\end{figure}


\section{{\nameM}: Metadata-private message store}\label{sec:pong}

\nameM{} is a message store that allows users to drop messages and retrieve them later, with the core goal of ensuring relationship unobservability~\cite{terminologyunoberser}. This means that an attacker cannot link the drop (write) and retrieval (read) operations for any given message.

A straightforward key-value store inside an enclave is \textit{not private}, as access patterns could reveal when a message was stored. Applying generic oblivious RAM (ORAM)~\cite{GoldreichOstrovsky1996Software,Goldreich1987Towards,optoramajacm,asharov2023futorama,patel2018panoramafocs,Stefanov2013PathORAM,Ren2015ConstantsCount} to hide all access patterns is \textit{not efficient} for instant messaging. ORAMs are designed to obscure both reads and writes under random access assumptions, which is an overkill in \nameM{}, where messaging dynamics naturally separate these operations.

Instead, \nameM{} leverages oblivious hash tables (OHT)~\cite{chan2017oblivioushashing,patel2018panoramafocs,asharov2023futorama}, a primitive commonly used in various ORAM constructions, as an efficient alternative for our purpose.
Building upon OHTs with customized optimizations, \nameM{} achieves unlinkability of messages while maintaining low latency.




\subsection{Oblivious Hash Table: The Building Block and Starting Point}

Oblivious hash table is a query-efficient hash table structure that ensures oblivious access of non-recurrent queries~\cite{chan2017oblivioushashing,patel2018panoramafocs}. 
We follow the oblivious hashing scheme defined in \cite{chan2017oblivioushashing}, simulated by an ideal hash functionality $\mathcal{F}_{\mathsf{HT}}$ as follows:

\begin{itemize}
    \item $\perp\gets\mathcal{F}_{\mathsf{HT}}.\texttt{Build}(\textsf{W})$: takes input of a set of real and dummy key-value pairs $\textsf{W} = \{(k_i,v_i)\| \textsf{dummy} \}_{i\in[n]} $, and builds the items into a hash table. This process is valid if any two non-dummy items' keys are distinct. 
    \item $v \gets \mathcal{F}_{\mathsf{HT}}.\texttt{Lookup}(k)$: takes input of a key $k$, and performs lookup functions on the hash table. If $k$ is dummy or $k \notin \textsf{W}$, then it will return $v = \perp$.
    
\end{itemize}

{\nameM} follows a black-box usage of readily available oblivious hash tables, and chooses the two-tier hash table design~\cite{chan2017oblivioushashing}. This design is noted for its simplicity and effectiveness with regard to implementation~\cite{DautermanFDCP21}. 
In this design, key-value pairs are distributed into buckets across two hash tables, $h_1$ and $h_2$, based on distinct keys. Items are primarily stored in $h_1$, with overflow items placed in $h_2$ as needed. This two-tier approach reduces the bucket size compared to a single-tier hash table.
The \textit{build} process involves four rounds of oblivious sorting to randomize the input set, disassociate items from their original order, and place them into appropriate buckets based on their keys. 
The \textit{lookup} operation computes bucket locations $\langle h_1(k), h_2(k) \rangle$ and performs a linear scan within these two buckets, achieving an $O(1)$ complexity for each query.

\noindent\textbf{Warmup: Unlinking writes and reads with oblivious hash table.}
Let us start with a simple example where we unlink write and read operations for one batch of messages $\mathsf{M} = \{k_i, m_i\}_{i \in [n]}$. 
By invoking $\mathcal{F}_{\mathsf{HT}}.\texttt{Build}$, we construct a hash table $\mathsf{OBin}$, where all messages are securely stored with their locations obscured. Once built, $\mathsf{OBin}$ can be used for future retrieval requests.
Oblivious hash tables require that real queries must be non-recurrent to maintain obliviousness. This requirement is naturally satisfied in our scenario, as the keys (tokens) used for message storage are uniquely managed by the \nameN{} system. As in \nameN, users always derive unique tokens for their messages as follows (Figure~\ref{fig:clientfunccode}, Line 24), 
\begin{lstlisting}[numbers=none,basewidth=0.5em]
    token = PRF(sk, myID + buddy.ID + str(counter)).
\end{lstlisting}
Since each token is guaranteed to appear only once, future lookup operations on $\mathsf{OBin}$ inherently preserve obliviousness.

This basic example demonstrates how OHT delinks writes from reads within a single round. However, as messages may not be immediately retrieved, $\mathsf{OBin}$ must persist for a reasonable period. When new batches of messages arrive, they must either be mixed with existing messages or handled through alternative methods.

\noindent\textbf{Append-based storage and optimizations.}
Since OHTs are inherently static structures, global mixing requires rebuilding the entire table, which is a costly operation. Instead, we take an append-based approach by creating new and small $\mathsf{OBin}$s for incoming messages while keeping previous tables intact. This allows retrieval operations to scan linearly across multiple OHTs, leveraging the efficiency of OHT lookup operations.
However, an unbounded increase in the number of $\mathsf{OBin}$s would result in longer linear scans and higher retrieval costs. To mitigate this, we introduce optimization strategies to control the growth rate of storage structures. 

\begin{figure}[t]
    \centering
    \includegraphics[width=0.82\linewidth]{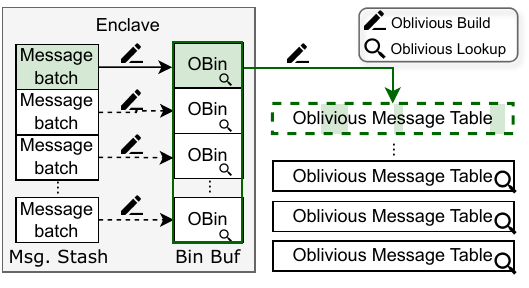}
    \vspace{-2mm}
    \caption{Storage structure of \nameM{}. }
    \label{fig:pong}
\end{figure}

Finally, we complete the whole picture of \nameM{}'s structure (Figure~\ref{fig:pong}).
It comprises a message stash, a bin buffer $\mathsf{Buf}$ with oblivious bins ($\mathsf{OBin}$), and $\mathsf{\textbf{T}}$, a set of large oblivious message tables ($\mathsf{OMT}$). Both $\mathsf{OBin}$ and $\mathsf{OMT}$ are oblivious hash tables of different sizes.

Upon arrival, a message batch is stored in the message stash and simultaneously constructed into a read-only $\mathsf{OBin}$. Periodically, $\mathsf{OBin}$s are merged into larger $\mathsf{OMT}$s to minimize lookup scans. 
For retrieval, each query invokes a lookup function on all oblivious hash tables ($\mathsf{OBin}$s and $\mathsf{OMT}$s).

\subsection{Storing Messages on \nameM{}}

Upon receiving a message (write query) batch $\mathsf{W} = \{k_i, v_i\}_{i \in[n]}$, the enclave first creates an oblivious message bin $\mathsf{OBin}$ by calling \texttt{OHT.Build}($\mathsf{W}$)\footnote{Due to space interest, the protocols are formally presented in Figure~\ref{fig:SimHom} in Appendix~\ref{app:proof}.}. %
Once the bin is built, this batch of messages is ready for future read access.
\nameM{} constructs a new $\mathsf{OBin}$ for each incoming batch and performs a linear scan across all existing bins for each read query. This structure effectively ensures the unlinking of write and read operations. However, for efficiency reasons, we aim to build larger oblivious bins that can store more data\footnote{Note that the size of the OHT does not affect lookup efficiency.}, reducing the total number of $\mathsf{OBin}$s.

The dilemma is that building individual large bin costs more time, e.g., superlinear in the total number of real items, which would incur delays for the newly incoming batches to become available for next query. 
Therefore, our optimization solution is two-fold: 1) letting each $\mathsf{OBin}$ contain more batches and 2) merging multiple $\mathsf{OBin}$s into a larger oblivious hash table.

\noindent\textbf{Optimization \#1: Reducing the number of $\mathsf{\textbf{OBin}}$s.} 
Inspired by hierarchical ORAMs \cite{GoldreichOstrovsky1996Software,optoramajacm,asharov2023futorama,patel2018panoramafocs}, \nameM{} periodically merges $k$ smaller bins into a larger bin, but in a simpler way. 
Specifically, over $k$ rounds, \nameM{} builds $k-1$ temporary $\mathsf{OBin}$s, each containing one batch of messages, and finally constructs a larger $\mathsf{OBin}$ with $k$ batches. This reduces retrieval query costs by a factor of $k$. 
However, constructing larger $\mathsf{OBin}$ (i.e., the \texttt{OHT.Build} function) is time-consuming and leads to inefficient worst-case write response time~\cite{asharov2023oramlog}. Therefore, we set $k$ to relatively small values (i.e., $k = 4$).

\noindent\textbf{Optimization \#2: Building larger oblivious message table on the background.} 
To further reduce the number without affecting the write rate, we propose a strategy similar to deamortization~\cite{OstrovskyS97,chan2017oblivioushashing,asharov2023oramlog}, which distributes the computational load of large oblivious hash tables evenly to ensure a consistent and low response time. 
The challenge lies in facilitating new read and write operations while some oblivious hash tables are being reconstructed. 

We leverage the read-only nature of bin accesses: by maintaining read-only copies of bins during their reconstruction, we ensure uninterrupted read and write operations. Once larger oblivious hash tables are built, the older, smaller bins are replaced and evicted from memory, enhancing read efficiency without compromising write response times.
Specifically, we keep the copies of $\mathsf{OBin}$s while merging them into a larger oblivious table. 
This way, the write query response time only involves a single \texttt{OHT.Build} on a small batch of messages.

\noindent\textbf{Optimization \#3: Keeping original copies.} 
To avoid expensive extraction of $\mathsf{OBin}$s during the construction of $\mathsf{OMT}$, we keep copies of original message data upon arrival, stored in a message stash. 
The message stash preserves the original messages $\mathsf{W}$, along with the bin index $idx$. Once the stash reaches its capacity of $m$ batches of messages, it calls a merge operation on all messages within the stash. 
The stash can be stored either in the in-enclave memory or the external memory, depending on available capacity.

Maintaining copies of the original messages requires additional memory costs. However, merging the oblivious bins directly would necessitate extracting real data from the bins first, incurring an oblivious compaction of $O(n\log n)$ time complexity. Since $\mathsf{OBin}$ will padded to be $10\times$ larger than the original real items, the extraction process would be notably slow. Thus, by prioritizing time over space, we manage to maintain an efficient reconstruction time.

\noindent\textbf{\textit{Remark}.} 
Theoretically, \nameM{}'s storage  can grow indefinitely given unlimited memory. However, this would result in increased lookup times. To balance usability and efficiency, \nameM{} maintains a fixed number of $N$ message batches, and expired messages are removed from storage.
This design choice means that if a client remains offline for an extended period, their unfetched messages may expire. This approach aligns with common practices in modern instant messaging applications. For example, messages in WeChat expire after three days~\cite{wechathelp}.

\subsection{Retrieving Messages from \nameM{}}

For each read request $\{k_i, v_i\}$ within the read batch $\mathsf{R}$, it iterates through every $\mathsf{OBin}$ in $\mathsf{Buf}$ and every $\mathsf{OMT}$ in $\mathsf{\textbf{T}}$. In each iteration, it conducts a lookup operation on the oblivious hash table. If the requested item is found, the system proceeds to perform dummy accesses on the remaining layers until completing all iterations.

To perform each read request, the system executes $|\mathsf{Buf}| + |\tilde{\textbf{T}}| = O(\epsilon N)$ times of \texttt{OHT.Lookup} operations. We have experimentally found that $\epsilon = 1/\sqrt{N/k}$ can optimize the read request running time (\S\ref{sec:micro}). 
Each \texttt{OHT.Lookup} incurs a constant cost, involving identifying the relevant bucket within the oblivious hash table and then carrying out a linear scan within the buckets for key comparison. This scan utilizes a series of register-level atomic oblivious operations, i.e., oblivious comparison and oblivious choose.


\section{Horizontal scalability}\label{sec:HSarchi}
Horizontal scalability allows a system to accommodate an increasing number of users by adding more servers. 
Horizontal scalability is crucial not only for performance but also for maintaining individual privacy in anonymous systems~\cite{sosp17Stadium,sosp17atom,osdi18karaoke,nsdi23boomerang, Anonymity06Dingledine}.

\name{} employs a readily available two-layer horizontal scaling architecture supported by hardware enclaves~\cite{nsdi23boomerang,DautermanFDCP21}. This architecture consists of entry nodes that perform batch assignment and distribution, and backend nodes that handle further processing of sub-batches. The system achieves scalability by evenly distributing workloads across multiple servers. The entry nodes function as load balancers, assigning queries to backend nodes based on unique identifiers like keys~\cite{DautermanFDCP21} or labels~\cite{nsdi23boomerang}. Through the use of max padding calculated via variants of balls-to-bins analysis, it is ensured that the output structure on each entry node is independent of the input content, with a marginal cost for padding.
Both \nameN{} and \nameM{} are designed to be adaptable to this structure, with necessary adjustments for batch distribution on the entry nodes. 


\noindent\textbf{Making \nameN{} horizontal scalable.}\label{sec:pinghorizontal}
To integrate \nameN{} into this two-layer architecture, we utilize carrier labels to distribute the workload among different backend nodes for global aggregation. This ensures that all notification vectors for the same client are directed to the same backend node, regardless of the entry node receiving the initial notification.

The entry nodes generate oblivious sub-batches by padding them to a predetermined upper bound, consistent with the balls-into-bins model, which requires labels to be random and distinct to avoid overflow and ensure correct distribution across backend nodes~\cite{nsdi23boomerang}.

To comply with this statistical model, additional operations on carrier labels are performed at the entry nodes. Due to significant potential variance in label occurrences, we employ a basic secure balls-into-bins model~\cite{DautermanFDCP21} rather than the weighted model from Boomerang~\cite{nsdi23boomerang}. In this model, entry nodes aggregate notification vectors to ensure each label appears only once, by adding modifications on the labels in Step 4 in Figure~\ref{fig:aggrcode} as follows:
\begin{lstlisting}[numbers=none, xleftmargin=-15pt, framexleftmargin=0pt, frame=none, aboveskip = 5pt, belowskip=5pt,basewidth=0.5em]
    Prev(pkt).label = OblChoose(is_rep, random(), Prev(pkt).label)
    prev_pkt.is_dummy = is_rep
\end{lstlisting}
This aggregation involves a linear scan, and to optimize time we can assign a backend node number to each packet during this process.

Next, the bins are assigned to the corresponding backend nodes according to the oblivious bin assignment algorithm~\cite{DautermanFDCP21,chan2017oblivioushashing}: 
\noindent
\begin{enumerate}[leftmargin=*,noitemsep,topsep=0pt,parsep=0pt,partopsep=0pt]
    \item Append $B*Z$ filler messages, tagged with $\mathsf{dummy}$, at the end of the sequence, where $B$ represents the number of backend nodes, and $Z$ is the derived upper bound.
    \item Obliviously sort the sequence by backend node ID (bin number) and dummy tag, ensuring that real data precedes dummy data for the same IDs.
    \item Perform a linear scan across the entire sequence, marking the first $Z$ messages in each bin for further processing, and tagging any excess messages for deletion.
    \item Employ an oblivious compaction to remove messages marked for deletion. Finally, distribute the sorted sub-batches to the respective backend nodes.
\end{enumerate}

Gathering all sub-batches, backend nodes then merge and process these sub-batches through the oblivious aggregation function (\S\ref{sec:aggregation}). This process only modifies packet payloads to avoid access pattern leakage. After processing, the packets are sent back to their originating entry node. 
The concluding steps of this process mirror those of the digest generation phase (\S\ref{sec:generatingdigest}). 

\noindent\textbf{Making \nameM{} horizontal scalable.}
The adaptation for \nameM{} is more straightforward due to the distinct keys appearing in queries. Write and read requests are divided into sub-batches based on their tokens at the entry node, following a similar oblivious batch assignment process as \nameN{}. This method allows messages to be stored across distributed storage (backend nodes) and read requests to be correctly routed to the appropriate backend nodes based on matching tokens.

\section{Security analysis}\label{sec:security}

Since confidentiality is inherently maintained in \name{} under hardware trust, the primary focus is on analyzing whether 1) observable interactions between users and servers, and 2) access patterns within enclaves could potentially allow a powerful global attacker to deduce whether two users are real communicating buddies. 
With the oblivious primitives and our algorithmic enforcement on traffic uniformity, we demonstrate that the possibility of such inferences is negligible.

\begin{myTheorem}(Informal)\label{them:sec}
\name{} achieves communication trace indistinguishability.
\end{myTheorem}

\emph{Proof~sketch}. 
Communication among users in \name{} can be formalized as multi-round interactions among a subset of users. 
We begin by establishing the indistinguishability of the interactions within \textit{a single round}, and then extend to \textit{multiple round} cases.

Each round of interaction is divided into two parts: message notification and message transmission (both sending and receiving), managed by \nameN{} and \nameM{} respectively. 
During the notification phase, the \nameN{} server prepares a set of carrier messages corresponding to the number of receivers. As notification packets from senders arrive, those vectors with the same carrier label are grouped and aggregated into a single message. 
This step effectively transits all notification for each user into one carrier message.
Since the packets are obliviously sorted and aggregated, the real mapping of the notifications to users is protected.
The enclave then obliviously compacts the sequence, extracting only the carrier messages, which are sent back to users. Therefore, from the perspective of attackers, they can only observe that every user sends and receives a packet to and from the \nameN{} servers. The relationships between these messages are obscured by the oblivious operations within  enclaves, rendering indistinguishable communication trace in \nameN{}.

Next, we show \nameM{} ensures the unlinkability of write and read requests across rounds, thus preserving the identity of sender and receiver of each message. From the message standpoint, each message from a sender is stored in an oblivious message bin along with a batch of write requests~\cite{chan2017oblivioushashing}. The \texttt{OHT.Build} function protects the position of each message within the bin. These bins are then placed in a buffer and later compiled into a larger message table via an oblivious build function. Although an attacker may deduce the bin (or table) in which a message is stored, the exact position within these containers remains concealed. 
When a user retrieves a message, it performs a linear-scan lookup of all bins and tables. Note that, OHT ensures that the lookup function will not leak any information as long as the non-dummy lookup queries are non-recurrent~\cite{chan2017oblivioushashing,optoramajacm}.
Combined with this linear scan approach, the retrieved messages by a read request are equally likely to be any of the messages stored on the storage system. Thus, the communication trace during the message transmission phase remains indistinguishable.

We then extend to scenarios where an attacker has knowledge of users' online/offline status across multiple rounds. 
\name{} requires that all users uniformly engage in sending and receiving messages (\S\ref{sec:schedule}), irrespective of real communication behavior. Consequently, the intersected online presence of two users does not provide sufficient information to confirm whether they are communicating buddies. This is because their online or idle status, which are independent of real communication activities, are not disclosed to attackers. Hence, interaction status observed in each round cannot be cumulatively used to distinguish communication traces across rounds. This completes the proof sketch. 

The formal security analysis is provided in Appendix~\ref{app:proof}.



\section{Implementation and evaluation}\label{sec:implementation} \label{sec:evaluation}

\subsection{Evaluation Overview}
Our evaluations are structured into two primary components: an empirical analysis of the notify-before-retrieval framework using real-world datasets and a performance evaluation of the PingPong prototype. These evaluations address the following questions:

\noindent\textbullet~ Why is the notify-before-retrieval framework more efficient than the dial-before-conversation framework on real-world messaging datasets? (Results are presented in Appendix~\ref{sec:evaluation_framework}.)

\noindent\textbullet~ How do \nameN{} and \nameM{} perform under varying workload, and what is the integrated performance?

Below are some highlighted results:

\noindent\textbullet~The notify-before-retrieval framework achieves $7.5\times$ less average waiting time on the real-world instant messaging workload. 

\noindent\textbullet~ \nameN{} achieves sub-second (0.934s) 99th-latency for 50K concurrent clients, and \nameM{} throughput reaches 20K fetches/s over a total storage of $2^{29}$ messages.










\subsection{Implementation and Evaluation Settings}

The \name{} prototype is built in about 10,000 lines of C++ code.
The implementation is built upon the Intel SGX v2.21 framework, with Intel SGX DCAP Driver v1.41. For oblivious primitives, we use Intel's AVX-512 SIMD instructions and the library from XGBoost~\cite{sgboostres}.
Network communication among clients and servers is facilitated using gRPC v1.35, operating in an asynchronous RPC mode and secured with TLS.
We will opensource the \name{} prototype once accepted. 

\noindent\textbf{Implementation.}
We implement the horizontal scalable version of \name{} on Azure Cloud VMs of DCsv3 series~\cite{AzureVM}, equipped with the 3rd Generation Intel Xeon Scalable processors with Intel SGXv2 support~\cite{xeon3}.  
The \nameN{} server cluster includes 32 $\texttt{DC8ds\_v3}$ instances (8~vCPU, 64~GB of memory with 32 GB EPC memory therein), including 24 entry nodes and 8  backend nodes. 
The \nameM{} server cluster includes 32 $\texttt{DC8ds\_v3}$ instances, including 2 entry nodes and 30 storage nodes. 
We use one $\texttt{D32d\_v5}$ instance (32~vCPU, 128~GB of memory) for simulated clients, with each sending one RPC request at a time. 
The instances are run in the same data center to save on bandwidth dollar costs. To compensate for the lack of inter-data center latency and simulate realistic network conditions, we introduced an additional 100 ms round-trip latency using the Linux $\mathsf{tc}$ command.
Results are averaged over 20 iterations for each experiment.

\begin{table}[t]
\centering
\small
\begin{tabular}{@{} lp{1.2em}p{1.2em}p{1.2em}p{1.2em}p{1.2em}p{1.2em} @{\hspace{1.3em}} p{1.2em} @{}}

& \rotatebox{30}{Pung~\cite{osdi16pung}}& \rotatebox{30}{Karaoke~\cite{osdi18karaoke}} & \rotatebox{30}{XRD~\cite{nsdi20XRD}} &  \rotatebox{30}{Boomerang~\cite{nsdi23boomerang}}&\rotatebox{30}{Groove~\cite{osdi22Groove}} & \rotatebox{30}{DPIR~\cite{DBLP:conf/ccs/ToveyWG24}}  &\rotatebox{30}{\textbf{PingPong}} \\
\toprule
Trust assumption      & \cellcolor{checkmarkGreen}Zero &\cellcolor{checkmarkGreen} Frac. & \cellcolor{checkmarkGreen}Frac.  & \cellcolor{checkmarkGreen}HW & \cellcolor{checkmarkGreen}Frac. & \cellcolor{checkmarkGreen}Zero & \cellcolor{checkmarkGreen}HW \\
Privacy strength  & \cellcolor{checkmarkGreen}C&\cellcolor{crossRed} DP & \cellcolor{checkmarkGreen}C & \cellcolor{checkmarkGreen}C &\cellcolor{crossRed} DP & \cellcolor{checkmarkGreen}C & \cellcolor{checkmarkGreen}C \\
No dialing      & \cellcolor{crossRed}\ding{55}& \cellcolor{crossRed}\ding{55} & \cellcolor{crossRed}\ding{55}  & \cellcolor{crossRed}\ding{55} & \cellcolor{checkmarkGreen}\ding{51} & \cellcolor{crossRed}\ding{55} & \cellcolor{checkmarkGreen}\ding{51} \\
%
%
%
Communication efficient & \cellcolor{crossRed}{\ding{55}}&\cellcolor{checkmarkGreen}\ding{51}&  \cellcolor{crossRed}{\ding{55}}  & \cellcolor{checkmarkGreen}\ding{51} &   \cellcolor{crossRed}{\ding{55}} & \cellcolor{crossRed}{\ding{55}} & \cellcolor{checkmarkGreen}\ding{51} \\
\bottomrule
\end{tabular}
\caption{Comparison of several representative metadata-private messaging systems based on trust assumption (\underline{Zero}, \underline{Frac}tional, or \underline{H}ard\underline{w}are trust), privacy guarantees (\underline{C}ryptographic privacy or \underline{D}ifferential \underline{P}rivacy), usability (whether need dialing), and communication efficiency.}\label{tab:baselines}
\end{table}

We set the maximum friend size as 512 by default. The notification packet is set to 256~Bytes with a 256-bit label. The retrieval token is of 64~bits. 
The message content is of 256~Bytes by default.  
We set the batch size $B$ according to~\cite{DautermanFDCP21}, with the security parameter $\lambda = 128$.
For the oblivious hash table, we set the security parameters $\lambda = 128$ and  $\epsilon = 0.75$, therefore the bucket size $Z$ is 17. 
The maximum batch size $c$ is a parameter related to the total number of clients. 
Each $\mathsf{OBin}$ contains $k = 4$ batches.
The maximum storage indicator $N$ is set to 8,640 by default.


\subsection{End-to-end Performance of \name{}}
Our evaluation measures the end-to-end latency and communication cost of \name{} across a total client base ranging from 10K to 100K. In each round, every client sends and receives one message, along with a notification packet.

\noindent\textbf{Selection of baseline.} 
Choosing an appropriate baseline is essential for a fair and meaningful comparison. Table~\ref{tab:baselines} lists several potential baselines. 
To the best of our knowledge, Groove~\cite{osdi22Groove} is the most recent system supporting dialing-free messaging. However, it relies on differential privacy and is not open-sourced, limiting its suitability as a direct baseline.
On the other hand, Boomerang~\cite{nsdi23boomerang} is a recent system that matches \name{} in terms of privacy guarantees, trust assumptions, and communication efficiency. However, Boomerang requires dialing by design, which contrasts with \name{}'s dialing-free usability goal. Thus, while Boomerang is a strong candidate, it requires adaptation to align with \name{}'s goals.

To create a baseline that satisfies both dialing-free messaging and cryptographic privacy, we modified Groove by replacing its differential privacy mixnet backend with Boomerang's mixnet. We denote this adapted baseline as Boomerang-BF. Specifically, Boomerang-BF 1) leverages Boomerang's enclave mixnet for metadata-private message exchange, and 2) adopts Groove's strategy of exhaustive enumeration of message exchange circuits to achieve dialing-free functionality. For the implementation,  we assign 24 $\texttt{DC8ds\_v3}$ instances as entry nodes and 8 instances as Boomerang nodes, following the optimal configuration reported in~\cite{nsdi23boomerang}.
This hybrid approach combines the strengths of Groove and Boomerang, ensuring a robust and comparable baseline for evaluating \name{}.

\begin{figure}[!t]
    \centering
    \includegraphics[width=0.9\linewidth]{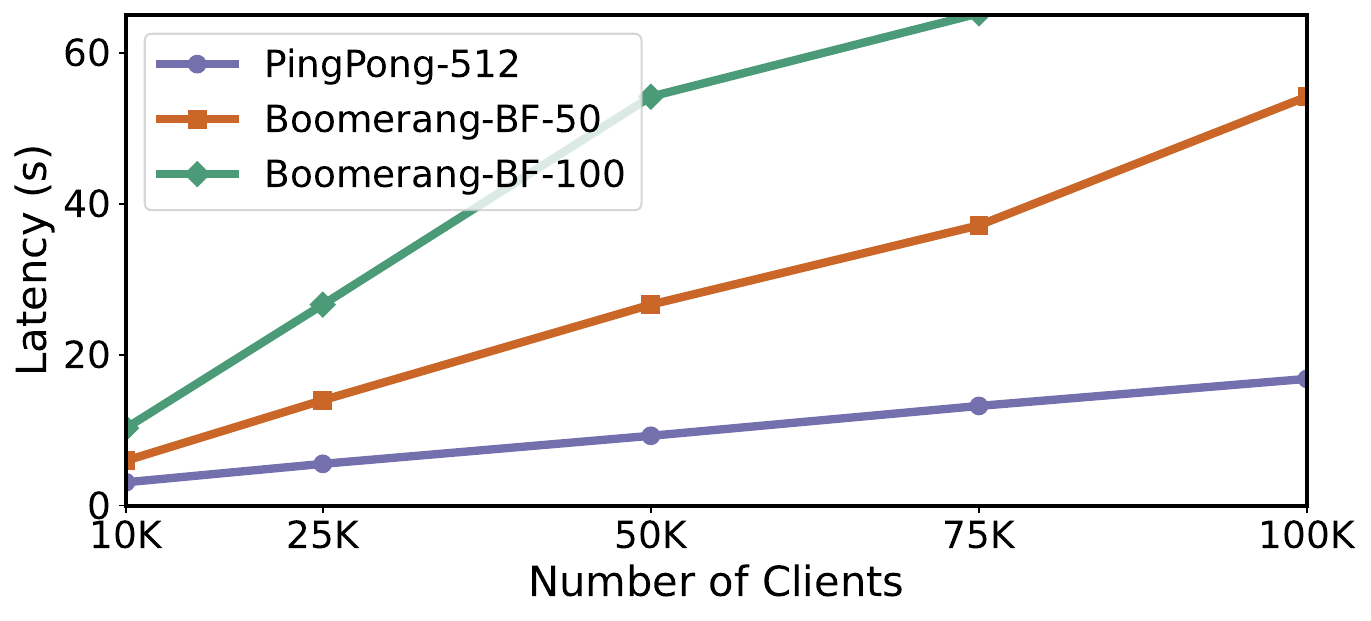}
    \vspace{-2mm}
    \caption{End-to-end 99th-latency of \name{} and Boomerang~\cite{nsdi23boomerang} (an adapted version that supports uncoordinated conversations with Groove's strategy~\cite{osdi22Groove}) with a varying number of total clients, each having 512, 50, and 100 MaxFriends, respectively.}
    \label{fig:ppe2eLatency}
\end{figure}

\noindent\textbf{Latency.}
Figure~\ref{fig:ppe2eLatency} compares the end-to-end 99th latency of \name{} and Boomerang-BF. With a client base of 100K, \name{} completes a full messaging cycle in just 16.80 seconds, significantly outperforming Boomerang-BF, which takes 54 seconds. Notably, \name{} achieves this efficiency while supporting a friend list size approximately 10 times larger than Boomerang-BF-50.

This improvement is primarily due to \name{} avoiding the maximum padding of friend sizes. With much fewer items, the shuffles and oblivious sorts in \name{} are considerably more lightweight than the brute-force approach employed by Boomerang-BF, which exhibits linear circuit growth with increasing friend counts.

\begin{figure}[!t]
    \centering
    \includegraphics[width=1.05\linewidth]{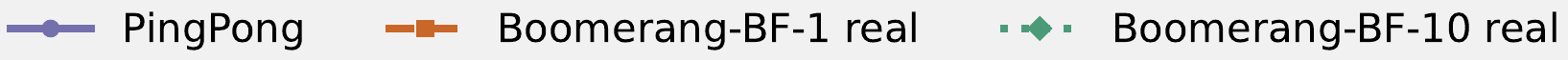}
    \subfigure[MaxFriends vs. Goodput]{
    \begin{minipage}[t]{0.475\linewidth}
        \centering
        \includegraphics[width=0.96\linewidth]{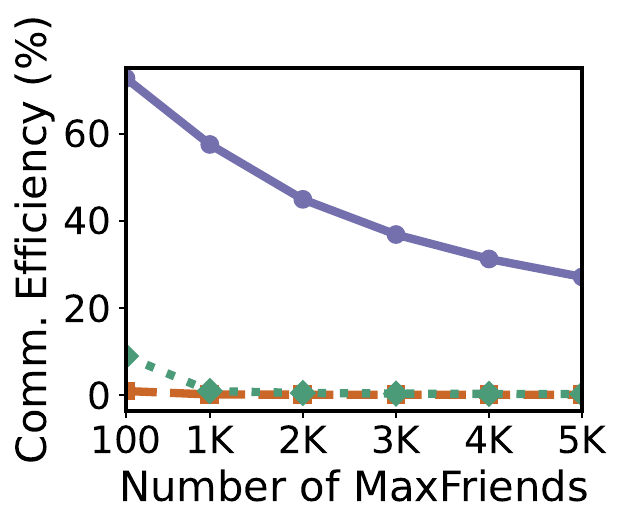}
        
        \label{fig:commcostratio}
    \end{minipage}}
    \subfigure[MaxFriends vs. Concrete cost]{
    \begin{minipage}[t]{0.475\linewidth}
        \centering 
        \includegraphics[width=0.96\linewidth]{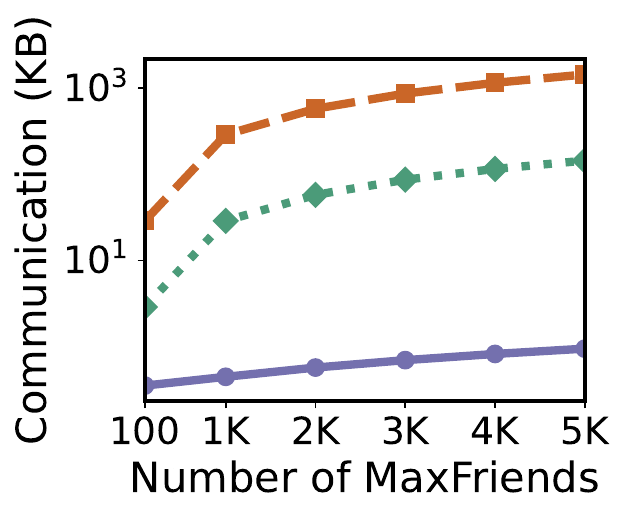}\label{fig:commcost} 
    \end{minipage}}
 
    \caption{Communication efficiency of \name. 
    ``Boomerang-BF-1/10 real'' means that every client receives 1 or 10 real messages in this round.}
   \label{fig:comm}
\end{figure}

\noindent\textbf{Communication cost.} 
Figure~\ref{fig:comm} presents the communication cost comparison between \name{} and Groove strategy~\cite{osdi22Groove}. Groove incurs significant overhead due to the extensive circuit padding required, resulting in the total communication volume handled by proxy servers scaling linearly with the number of friends. For instance, sending a single 256-byte message per round results in the proxy transmitting 10,000 messages (about 2.56 MB) for a client with 5,000 maximum friends, even when only one real message is being exchanged.

\name{} employs a more efficient method by notification, involving only a single bit to indicate a friend's notification action. This reduces communication costs, which increase moderately with the maximum friend size. For 5,000 MaxFriends, the actual communication cost is reduced to 945 bytes per real 256-byte message. 


\subsection{Evaluation on \nameN{} and \nameM}

\noindent\textbf{Performance  of \nameN (Figure \ref{fig:pingoverall}).}
 We select Scalable PS~\cite{scalablePS} as the baseline for \nameN. Scalable PS is an advanced version of private signaling~\cite{PS22sec} and also under hardware trust. 
To make it metadata-private, we modified the client interactions to include both read and write requests, with the server processing them sequentially. 
We configured a single \texttt{DC8ds\_v3} instance to handle 1/32 of the requests managed by \nameN{}.

\nameN{} achieves latencies between 0.2s and 1.7s for client numbers from 10K to 100K, which surpasses the baseline system, particularly at higher client numbers. \nameN{}'s efficiency stems from its ability to batch requests, allowing for the simultaneous processing of multiple messages through oblivious aggregation techniques. While Scalable PS is effective with individual requests, its use of generic ORAM operations does not suit scenarios with intense batch queries.


 \begin{figure}[!t]
    \centering
    \subfigure[Build vs size]{
    \begin{minipage}[t]{0.3\linewidth}
        \centering \vspace{-3mm}
        \includegraphics[width=0.9\linewidth]{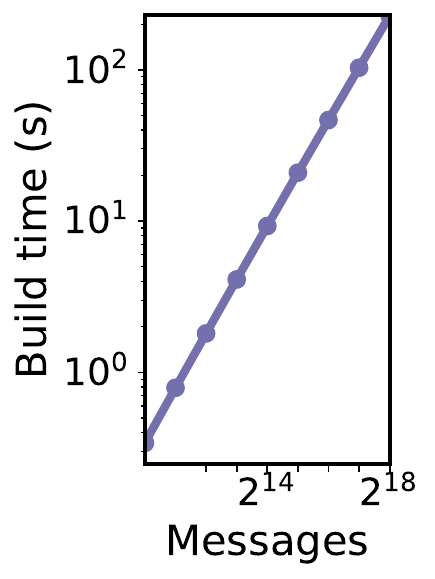}
        
        \label{fig:owritetime}
    \end{minipage}}
    \subfigure[Lookup vs layers]{
    \begin{minipage}[t]{0.3\linewidth}
        \centering \vspace{-3mm}
        \includegraphics[width=0.9\linewidth]{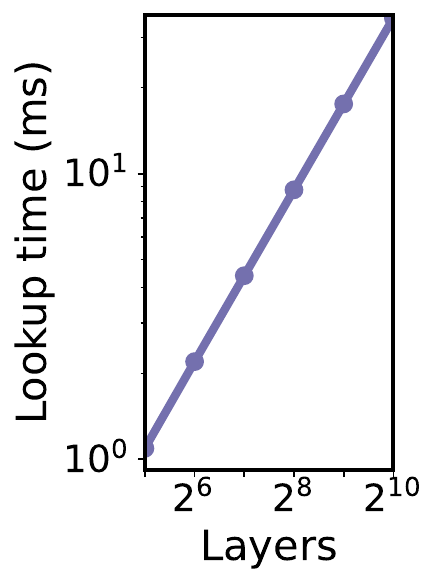}
        \label{fig:oreadtime}
    \end{minipage}}
        \subfigure[Lookup vs stash]{
    \begin{minipage}[t]{0.3\linewidth}
        \centering \vspace{-3mm}
        \includegraphics[width=0.9\linewidth]{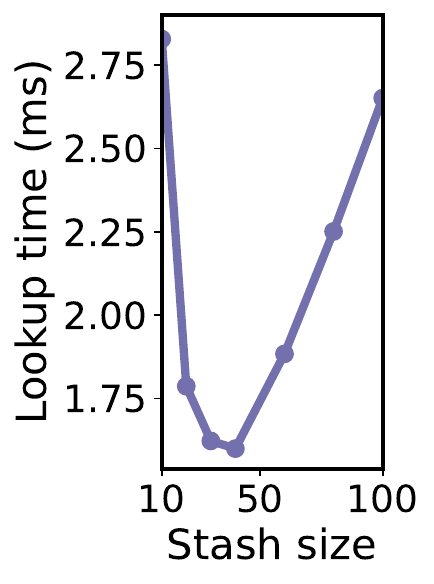}
        \label{fig:olayernum}
    \end{minipage}}
    \vspace{-3mm}
    \caption{Benchmark on oblivious hash tables.  }
    \label{fig:multi-dedup}
\end{figure}

\noindent\textbf{Performance  of \nameM (Figure~\ref{fig:pongoverall}).}
 The overall round latency involves processing a batch of write queries and read queries. 
When compared with the overall latency in \name{}, \nameM{} demonstrates a dominant share in response time. 
Since we carefully balance the write and read computational cost, they both grow linearly with the batch size. 
Note that the system is horizontally scalable, we can always add more servers to amortize the workload of queries. 


\subsection{Microbenchmarks}\label{sec:micro}
\begin{figure*}[!t]
  \begin{minipage}[t]{0.285\linewidth}
    \centering
    \includegraphics[width=0.98\linewidth]{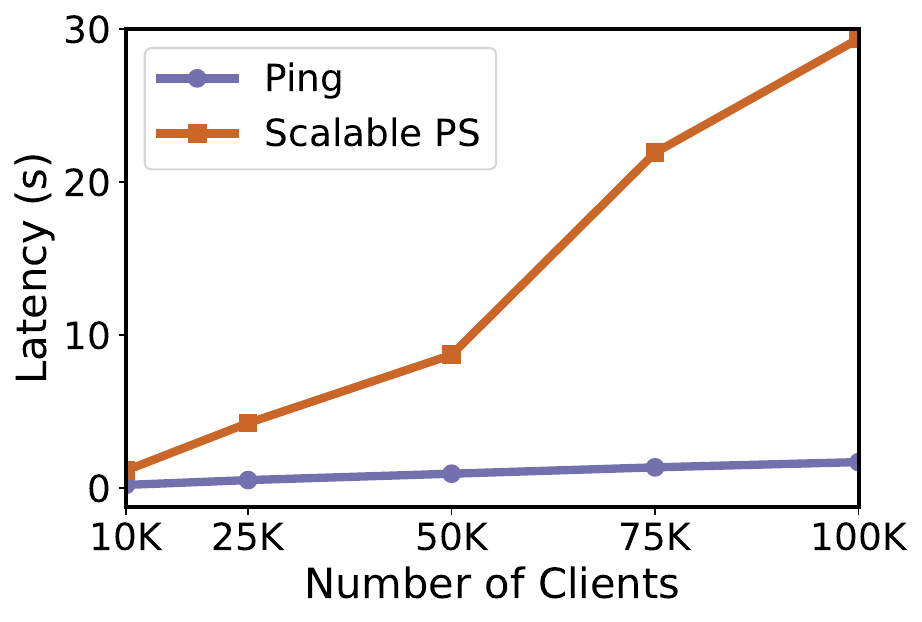}
    \caption{Comparison of \nameN and Scalable PS~\cite{scalablePS}.}
    \label{fig:pingoverall}
    \end{minipage}
   \hfill 
  \begin{minipage}[t]{0.285\linewidth}
    \centering
    \includegraphics[width=0.98\linewidth]{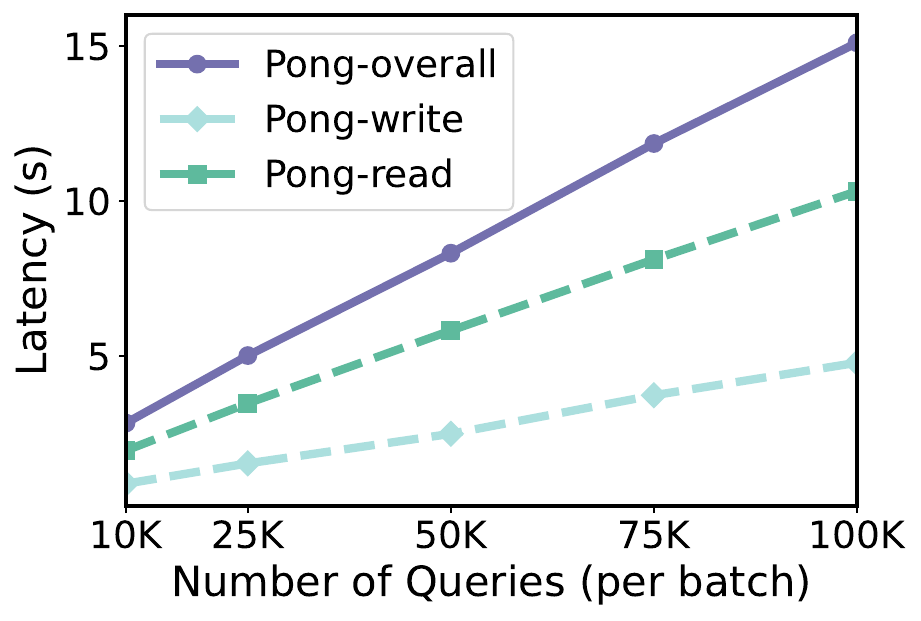}
    \caption{\nameM's latency  with different number of concurrent queries. }
    \label{fig:pongoverall}
    \end{minipage}
   \hfill 
   \begin{minipage}[t]{0.41\textwidth}
    \centering
    \includegraphics[width=0.98\textwidth]{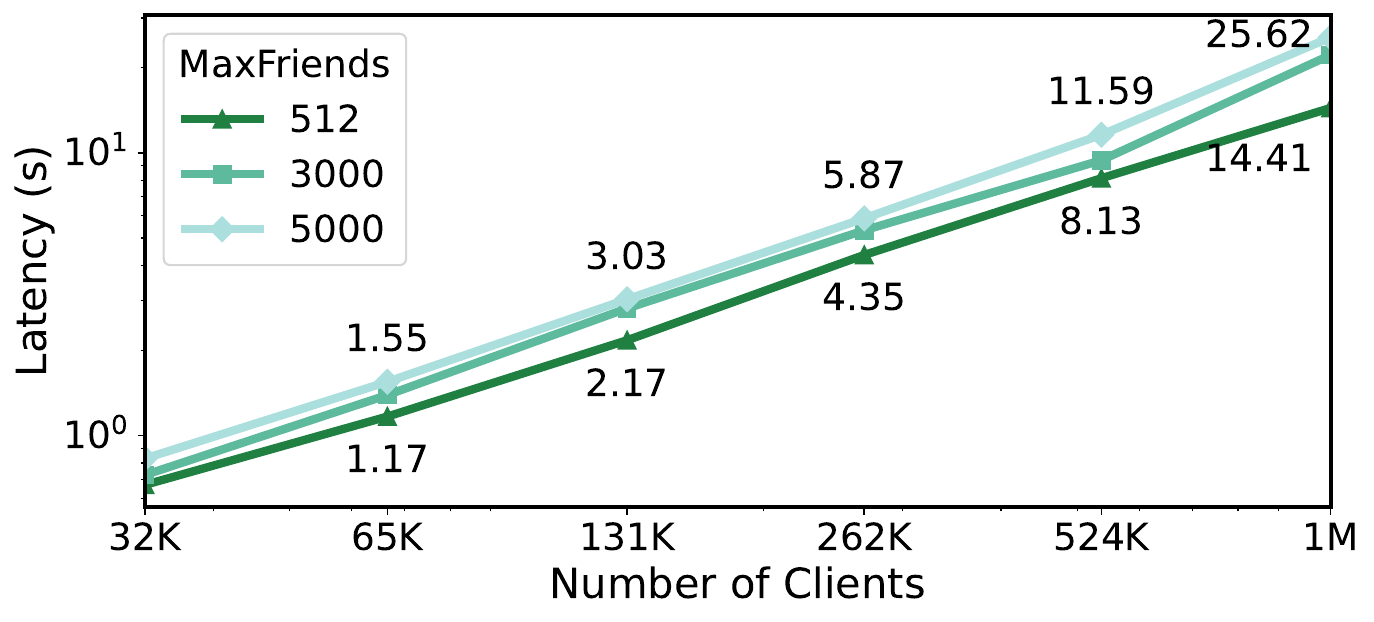}
    \caption{\nameN{}'s latency with a varying number of total clients and maximum friends.}
    \label{fig:pinglatencyfrdnum}
    \end{minipage}
   \hfill 
\end{figure*}

\noindent\textbf{Impact of MaxFriend (Figure~\ref{fig:pinglatencyfrdnum}).} 
To reflect real-world scenarios requiring more friends, 
we expanded \nameN{} to accommodate up to 5,000 friends and 1 million clients. As shown, the latency grows sublinearly with the maximum friend size.

\noindent\textbf{Oblivious hash table (Figure \ref{fig:multi-dedup}).}
The OHT is a critical component in \nameM{}, playing a key role in determining the parameters. Benchmark results for constructing an OHT with varying message counts and performing lookups across different numbers of OHT layers using linear scans are presented in Figures~\ref{fig:owritetime} and~\ref{fig:oreadtime}, respectively.
Since the batch size is relatively stable with the client size, the time required to build an $\mathsf{OBin}$ remains nearly constant. 
Thus, the main factors affecting performance are the number of $\mathsf{OBin}$s and $\mathsf{OMT}$s that need to be accessed during a read query.
Suppose an $\mathsf{OMT}$ contains $m$ $\mathsf{OBins}$. Given a fixed total batch capacity of $N$, the total number of OHTs is $m + N/km$. Consequently, it follows that when $m = \sqrt{N/k}$, the total number of lookups is optimized, as demonstrated in Figure~\ref{fig:olayernum}.


\noindent\textbf{Comparison with generic ORAMs (Table~\ref{tab:benchmark}).} 
For evaluating \nameM{}, we further use EnigMap~\cite{TinocoGS23}, a state-of-the-art enclave-based ORAM design, as the baseline. In
this experiment, both systems were implemented on a single
\texttt{DC32ds\_v3} instance.
For \nameM{}, each $\mathsf{OBin}$ contains 5,000 messages, and we set OMT size, $m$, according to the square root strategy. The benchmarks substantiate our hypothesis that for smaller-sized databases, a linear scan approach can outperform general-purpose ORAMs, despite the latter's theoretical optimality. Note that responding time for write queries in \nameM are only determined on the construction time of OBins. Meanwhile, OMTs can be built in parallel in the background, thereby not impacting the immediacy of write operations.


\begin{table}[!t]

\caption{Comparison of running time ($\mu$s) per read (search) / write (insertion) query for EnigMap and \nameM with different total entries in the storage. The batch size for \nameM{} is 5,000. }
\label{tab:benchmark}
\centering\small
\begin{tabular}{c|cc|cc}
\toprule
\multirow{2}{*}{\textbf{Total messages}} & \multicolumn{2}{c|}{\textbf{EnigMap}~\cite{TinocoGS23}} & \multicolumn{2}{c}{\textbf{\nameM{}}} \\ 
                                        & \textbf{Read} & \textbf{Write} & \textbf{Read} & \textbf{Write} \\ \midrule
$10^5$ & 2,125      & 4,337       & 388        & 832      \\ 
$10^6$ & 3,092      & 6,237       & 1,200      & 832      \\ 
$10^7$ & 4,053      & 8,207       & 3,431      & 832       \\ \bottomrule
\end{tabular}
\end{table}

\section{Related Work}\label{sec:relatedwork}

\noindent\textbf{Metadata-private communications.} Metadata-private communications have seen numerous advancements in recent years. From a technical perspective, these advancements can be broadly categorised into: 1) approaches that follow the mix-nets paradigm~\cite{DBLP:journals/cacm/Chaum81} with various security and privacy enhancements~\cite{sosp17atom, popets16riffle, nsdi20XRD, osdi18karaoke, sosp19yodel, osdi16alpenhorn, sosp17Stadium, sosp15vuvuzela, uss17loopix, DBLP:conf/sigcomm/BlondCCDM15, DBLP:conf/sigcomm/Le-BlondCZDBF13,nsdi23boomerang,osdi22Groove,ndss23trellis}; 2) approaches that follow the dining cryptographers network (DC-net)~\cite{Chaum88} and subsequent improvements on scalability and service resilience~\cite{DBLP:conf/ccs/Corrigan-GibbsF10,DBLP:conf/osdi/WolinskyCFJ12,DBLP:conf/uss/Corrigan-GibbsWF13,DBLP:conf/ccs/AbrahamPY20,nsdi22Spectrum_broadcast}; 3) more recent cryptographic proposals that utilize private information retrieval~\cite{osdi16pung, angel2018pir, osdi21Addra,DBLP:conf/ccs/ToveyWG24}, MPC~\cite{DBLP:conf/uss/AlexopoulosKT017} and distributed point function techniques~\cite{riposte15sp, uss21express,nsdi22Spectrum_broadcast,DBLP:conf/ccs/AbrahamPY20}, and fully homomorphic encryption~\cite{LiuTW23,OMR22cropto}. These systems generally strike a balance among security, e.g., cryptographic security~\cite{nsdi20XRD,uss21express} vs. differential privacy~\cite{sosp15vuvuzela,osdi18karaoke,osdi22Groove}, performance, e.g., horizontal scaling~\cite{nsdi20XRD,sosp17atom,nsdi23boomerang}, and trust assumptions, e.g., any-trust~\cite{sosp15vuvuzela} vs. fractional trust~\cite{nsdi20XRD}. For more detailed discussions, we refer to an excellent survey~\cite{cryptoeprint:2023/313}. 

\name{} relates to a performant subset of this large body of prior work, namely bi-directional metadata-private messaging systems~\cite{sosp15vuvuzela,nsdi20XRD,osdi18karaoke,sosp17Stadium,osdi16alpenhorn,osdi16pung,osdi21Addra,angel2018pir,DBLP:conf/uss/AlexopoulosKT017,nsdi23boomerang}. As mentioned earlier, these systems generally follow the ``dial-before-converse'' framework, which includes a resource-intensive dialing protocol and imposes usability limitations. \name{} overcome these limitations with a new ``notify-before-retrieval'' workflow, by integrating an metadata-private notification system and an metadata-private message store.

\noindent\textbf{Private signaling and oblivious message retrieval.} Private signaling (PS)~\cite{PS22sec, scalablePS} and oblivious message retrieval (OMR)~\cite{OMR22cropto, LiuTW23, DBLP:conf/ccs/JiaMK24} are two recently proposed primitives aimed at addressing the receiver-privacy problem, which is pertinent to but different from the problem \name{} tackles. While PS (based on secure enclave or MPC) and OMR (based on FHE) employ different technical constructions, both primitives strive to achieve the same privacy guarantee: a message cannot be linked to a receiver. Fuzzy Message Detection \cite{DBLP:conf/ccs/BeckLM021,DBLP:conf/fc/SeresPB22}, as an earlier notion than PS and OMR, shares the same goal of concealing the receiver-privacy problem, achieved through the use of false-positive fake messages.

We note that both PS and OMR are standalone protocols that focus solely on unlinkability of messages for a single receiver. This contrasts with end-to-end metadata-private communication systems like \name{}, which explicitly addresses the problem of relationship unobservability in settings involving a large number of users. The metadata privacy requirement in this context necessites the need for traffic uniformity, aka cover traffic, as part of the design space. This requirement has been witnessed by many recent advancements of full-fledged bidirectional private messaging systems~\cite{sosp15vuvuzela,nsdi20XRD,osdi18karaoke,sosp17Stadium,osdi16alpenhorn,osdi16pung,osdi21Addra,DBLP:conf/uss/AlexopoulosKT017,nsdi23boomerang}.

\noindent\textbf{Oblivious RAM.} ORAM is a fundamental concept in cryptography, designed to protect memory access patterns to ensure data security~\cite{GoldreichOstrovsky1996Software,Goldreich1987Towards}. 
The field of ORAM has evolved into several branches: 1) Hierarchical ORAMs: as originally conceptualized~\cite{GoldreichOstrovsky1996Software,Goldreich1987Towards}, with notable recent enhancements~\cite{optoramajacm,asharov2023futorama,patel2018panoramafocs,optoramaconf,chan2017oblivioushashing}, 2) tree-based ORAMs: exemplified by PathORAM~\cite{Stefanov2013PathORAM} and its variations~\cite{Ren2015ConstantsCount}, and 3) hardware-assisted ORAMs: utilizing secure enclaves for improved performance \cite{MishraPCCP18,TinocoGS23,SasyGF18} and high-throughput~\cite{DautermanFDCP21}.
Despite the effectiveness of ORAMs in protecting memory access patterns, direct application of these ORAMs as a backend for our message storage \nameM{} may be an over-kill. General-purpose ORAMs might introduce unnecessary computation operations, given our specific goal is only to unlink the writes and reads. Therefore, we made customized optimizations in \nameM{}  to fit our needs, which results in considerable performance improvement in \name{}.

\section{Conclusions}


Despite the recent progress of metadata-private communication systems, one major hurdle in almost all prior art is the need for coordination to initiate conversations, which can be expensive for both service adoption and operations. \name{} overcomes this limitation by implementing an alternative ``notify-before-retrieval'' workflow. It is integrated with one metadata-private notification system and one metadata-private message store, both leveraging hardware-assisted secure enclaves for performance and operating through a series of customized oblivious algorithms and protocols. 
The improved usability of \name{}, combined with its good performance and overall system goodput, marks a leap towards widespread adoption of metadata-private messaging in practice. 

\bibliographystyle{IEEEtran}
\bibliography{references}
\appendix

\subsection{Client Main Loop}\label{sec:cltloop}
\name{}'s components include: a set of clients, \nameN{} system, and  \nameM{} system. 
Upon establishing an online connection, clients engage in two  loops of interactions with the \nameN{} and \nameM{} servers as described below and illustrated in Figure~\ref{fig:clientloop}.

\noindent\textbf{Sending loop.}
The client begins by popping up a message from the message queue and generating a notification packet and a message packet, by invoking the \texttt{GenNotf} function for notification packets and the \texttt{GenMsg} function for message packets (\S\ref{sec:generatingnotf}). Once prepared, these packets are sent to the respective \nameN{} and \nameM{} servers.
Upon receiving a notification batch, \nameN servers obliviously aggregate the notifications and generate a notification digest for each recipient. 
Meanwhile, \nameM servers write the message packets to the oblivious  store. 


\noindent\textbf{Receiving loop.}
Upon receiving the notification digest, the user decodes the digest to retrieval tokens by referencing the local friend list, with \texttt{ParseNotf} function (\S\ref{sec:generatingnotf}). These tokens are then queued for scheduling (\S\ref{sec:schedule}).
The user then pops up a token from this queue and generates a retrieval request to \nameM{}. 
\begin{figure}[t]
\centering
\begin{tcolorbox}[
    enhanced,
    colback=white, 
    colframe=white, 
    boxrule=1pt, 
    left=-8pt, 
    right=1pt, 
    top=0pt, 
    bottom=0pt, 
  ]
\small\centering\textbf{------ Sending thread ------}
\begin{lstlisting}[basewidth=0.5em]
while True:
    buddy, content = GetInputFromQueue()
    # Generate notification and message packets
    notf_pkt = GenNotf(buddy)
    msg_pkt = GenMsg(buddy, content)
    # Send a packet to Ping
    try:
        client.SendNotf(notf_pkt)
    except Exception as errNotf:
        raise Exception("Notification Error") from errNotf
    # Send a packet to Pong
    try:
        client.SendMsg(msg_pkt)
    except Exception as errMsg:
        raise Exception("Message Error") from errMsg
\end{lstlisting}

\small\centering\textbf{------ Receiving thread ------}
\begin{lstlisting}[basewidth=0.5em]
while True:
    # Receive notification digests from Ping servers
    notf_pkt = client.RecvNotf()
    notf_vec = Decrypt(notf_pkt, PingKey)
    # Parse the notification into tokens
    msg_tokens = ParseNotf(notf_vec)
    # Add tokens to the token queue
    token_queue.push_tokens_to_queue(msg_tokens)
    # Get a token from the token queue
    token = token_queue.get_token_from_queue()
    # Generate a request packet
    pkt = Encrypt((token, []), PongKey)
    # Send the request packet to Pong servers
    response_pkt = client.SendReq(pkt)
    Buddy_ID, enc_data = Decrypt(response_pkt, PongKey)    
    # Lookup the secret key from FriendList
    buddy, exists = FriendList.map.get(Buddy_ID)
    if exists:
        msg = Decrypt(enc_data, buddy.sk)
\end{lstlisting}
\end{tcolorbox}
\vspace{-5mm}
  \caption{Client main loop.}
    \label{fig:clientloop}
\end{figure}

\subsection{Formal Security Analysis}\label{app:proof}

Following the proof sketch (\S\ref{sec:security}), we prove the security of \name{} modularly: we first prove that \nameM{} and \nameN{} achieve both communication trace indistinguishability in one communication round, and then show that this indistinguishability can be extended to multiple rounds when introducing the idle states in \name{}. 

\begin{myTheorem}\label{them:ping}
Given an oblivious assign algorithm $\mathsf{OblAssign}$, an oblivious choose algorithm $\mathsf{OblChoose}$, an oblivious sort algorithm $\mathsf{OSort}$, an oblivious compact algorithm $\mathsf{OCompact}$, and a secure encryption scheme $(\mathsf{Encrypt}$, $\mathsf{Decrypt})$, the message notification system \nameN{} achieves communication trace indistinguishability as defined in Definition~\ref{def:ctu} in every communication round.
\end{myTheorem}

\noindent\textbf{Proof.}
Intuitively, as all information in the notification is transmitted and operated in the encrypted form, to prove the indistinguishability, the key is to demonstrate that adversaries cannot distinguish the communication trace from memory access and communication patterns (i.e., interactions among clients and servers). In terms of communication patterns, as analysed in the proof sketch (\S\ref{sec:security}), \nameN{} makes sure that each client will send and receive exactly one messages to and from the servers. Furthermore, when adapted the two-layer architecture consisting of entry and backend nodes (Appendix \ref{sec:pinghorizontal}), the oblivious sub-batch generation protocol in \cite{nsdi23boomerang} ensures that the interactions among the servers and the communication pattern of the clients are indistinguishable even with active attackers.  
As a result, the attacker can only observe each online client (including both idle and true) sends and receives a message but cannot identify which two clients are engaged in communication with each other. Therefore, in this part, we mainly focus on the security of the access pattern.

To show the access patterns are indistinguishable, we construct a simulator algorithm for the oblivious aggregation and compaction step in Figure~\ref{fig:aggrcode} (which we denote as $\mathsf{OblAggregation}$) and demonstrate that this algorithm outputs a set of carrier messages that cannot be distinguished. 
In other words, the attacker cannot know which and how many input notification vectors are associated with the output carrier messages. 
As in Figure~\ref{fig:simaggrcode}, the $\mathsf{SimOblAggregation}$ algorithm takes as input an array of $\texttt{pkts}$ with the same size of the objects that will input to the original algorithm $\mathsf{OblAggregation}$, which is randomly generated by the simulator. Since the inputs of both algorithms have the same size and are encrypted, they are indistinguishable. With them as inputs, $\mathsf{OblAggregation}$ and $\mathsf{SimOblAggregation}$ first execute oblivious sort algorithm $\mathsf{OSort}$ and $\mathsf{SimOSort}$ respectively to sort packets (Line 1). By the security of the oblivious sort algorithm, it is clear that this operation produces indistinguishable memory access patterns. After that, They scan to process the packets as follows:
\begin{itemize}
    \item (Line 3) These lines check whether the neighboring packets have the same label by an oblivious algorithm $\mathsf{OEqual}$. Thus, the access patterns are indistinguishable. 
    \item (Line 4) These lines are identical in $\mathsf{OblAggregation}$ and $\mathsf{SimOblAggregation}$  which produce indistinguishable access patterns.
\item (Line 5-6) These lines proceed with the oblivious choose algorithm. The original $\mathsf{OblChoose}$ algorithm and the corresponding $\mathsf{SimOblChoose}$ work with indistinguishable memory access patterns.
\item (Line 7) These lines are identical in $\mathsf{OblAggregation}$ and $\mathsf{SimOblAggregation}$ and they will produce indistinguishable access patterns.
\end{itemize}

Finally, $\mathsf{OblAggregation}$ and $\mathsf{SimOblAggregation}$ execute oblivious compact algorithm $\mathsf{OCompact}$ and $\mathsf{SimOCompact}$ respectively to extract the carrier messages (Line 9). The indistinguishablity of this step is guaranteed by the oblivious compaction function. 

As seen above, memory access patterns in $\mathsf{OblAggregation}$ and $\mathsf{SimOblAggregation}$ are indistinguishable. Therefore, the attacker cannot identify the real and simulated aggregation algorithms. Combined with the restriction that they have the same size inputs, the output packets will be indistinguishable as well. Therefore, the message notification system \nameN{} achieves communication trace indistinguishability as defined in Definition~\ref{def:ctu} in one communication round.

\begin{figure}[t]
\begin{tcolorbox}[
    enhanced,
    colback=white, 
    colframe=white, 
    boxrule=1pt, 
    left=-10pt, 
    right=2pt, 
    top=2pt, 
    bottom=2pt, 
  ]
\underline{\textbf{Proc} $\texttt{SimOblAggregation}(\mathsf{pkts})$:} 

\begin{algorithmic}[1]\small
\State $\mathsf{pkts}.\texttt{SimOSort}(\mathsf{key} = \mathsf{pkts}.\mathsf{label}|| \mathsf{pkts}.\mathsf{is\_carrier})$
\For {$\mathsf{pkt}\in \mathsf{pkts}$}
\State $\mathsf{isRep} = \texttt{OEqual}(\mathsf{pkt.label}, \mathsf{Prev(pkt).label})$
\State $\mathsf{agg\_vec} = \mathsf{Prev(pkt).NotfVec} \vee \mathsf{pkt.NotfVec}$
\State $ \mathsf{pkt.NotfVec} = \texttt{SimOblChoose}(\mathsf{isRep}, \mathsf{agg\_vec}, \mathsf{pkt.NotfVec})$
\State $\mathsf{Prev(pkt).label} =\texttt{SimOblChoose}(\mathsf{isRep}, \text{random()}, \mathsf{pkt.label})$
\State	$\mathsf{Prev(pkt).is\_dummy} = \mathsf{isRep}$
\EndFor
\State $\mathsf{pkts}.\texttt{SimOCompact}(\mathsf{key} = \mathsf{is\_carrier})$
\end{algorithmic}

\end{tcolorbox}
\vspace{-3mm}
  \caption{Pseudocodes for $\texttt{SimOblAggregation}$.}
    \label{fig:simaggrcode}
\end{figure}

\begin{figure*}[!t]
\begin{tcolorbox}[
    enhanced,
    colback=white, 
    colframe=black, 
    boxrule=1pt, 
    left=2pt, 
    right=2pt, 
    top=2pt, 
    bottom=1pt, 
  ]
\begin{multicols}{2}

\underline{$\textsf{st} \leftarrow \texttt{Pong.OblWrite}(\mathsf{W}, \mathsf{Buf},\mathsf{Stash},  \mathsf{\textbf{T}},  \textsf{st})$:} 

\noindent\textbf{Input:} the write batch $\mathsf{W} = \{k_i, v_i\}_{i = 1...n\leq c}$, the bin buffer $\mathsf{Buf}$, the message stash $\mathsf{Stash}$, the set of oblivious message tables $\mathsf{\textbf{T}}$.
\begin{algorithmic}[1]

\State
\If {$st.tempBin < k-1$}
\State $\mathsf{OBin} := \texttt{OHT.Build}(\mathsf{W})$ \Comment{Create an oblivious bin}
\State $st.tempBin++$
\State Add $\mathsf{W}$ to $\mathsf{TempStash}$
\Else 
\State Remove $\{\mathsf{OBin}_i\}_{i = 1 ... k-1}$ from $\mathsf{Buf}$
\State $\mathsf{OBin} := \texttt{OHT.Build}(\mathsf{W}\cup\mathsf{tempStash})$ \Comment{Create a larger $\mathsf{OBin}$}
\State $st.tempBin = 0$
\EndIf
\State $idx\gets\mathsf{Buf}.\texttt{push}(\mathsf{OBin})$ \Comment{Add the bin to $\mathsf{Buf}$}
\State Add $(\mathsf{W}, idx)$ to $\mathsf{Stash}$
\If{$|\mathsf{Stash}| \geq m$}   
\State \Call{OblMerge}{$\mathsf{Stash}$} \Comment{Proceed on the background}
\EndIf
\Procedure{OblMerge}{$\mathsf{Stash}$}
\State Extract $\mathsf{Stash}$ to  $\{\mathsf{W}_i\}_{i = 1...m}, I = \{idx_i\}_{i = 1...m}$
\State $\mathsf{OMT} := \texttt{OHT.Build}(\bigcup^{m}_{i=1}W_i)$
\State Add $\mathsf{OMT}$ to $\mathsf{\textbf{T}}$
\State Remove $\{\mathsf{OBin}_i\}_{i \in I}$ from $\mathsf{Buf}$
\EndProcedure
\State Update $\textsf{st}$

\end{algorithmic}

\vspace{1pt}
\underline{$\mathsf{R'} \leftarrow \texttt{Pong.OblRead}( \mathsf{R}, \mathsf{Buf},  \mathsf{\textbf{T}},  \textsf{st})$:} 

\noindent\textbf{Input:} the read batch $\mathsf{R}= \{k_i, v_i\}_{i = 1...n}$, the bin buffer $\mathsf{Buf}$, the set of oblivious message tables $\mathsf{\textbf{T}}$.
\begin{algorithmic}[1]
\State
\For {$(k_i, v_i) \in \mathsf{R}$}
\State Initialize a global boolean tag $\textsf{found} = \textsf{false}$. 
\For{$\mathsf{OBin} \in \mathsf{Buf}$ } 
\State $k = \texttt{OblChoose}(\textsf{found}, k, \perp)$
\State $v' = \mathsf{OBin.}\texttt{Lookup}(k)$
\State $\textsf{OHTFound} = v'!= \perp $
\State $v = \texttt{OblChoose}(\textsf{OHTFound}, v', v)$
\State $\textsf{found} = \textsf{OHTFound} \vee \textsf{found}$
\EndFor  
\For{$\mathsf{OMT} \in \mathsf{\textbf{T}}$ } 
\State $k = \texttt{OblChoose}(\textsf{found}, \perp, k)$
\State $v' = \mathsf{OMT.}\texttt{Lookup}(k)$
\State $\textsf{OHTFound} = v'!= \perp $
\State $v = \texttt{OblChoose}(\textsf{OHTFound}, v', v)$
\State $\textsf{found} = \textsf{OHTFound} \vee \textsf{found}$
\EndFor  
\State  Append $(k, v)$ to $\mathsf{R'}$. 
\EndFor
\end{algorithmic}


\underline{$\textsf{st} \leftarrow \texttt{SimPong.OblWrite}(|\mathsf{W}|, \mathsf{Buf},\mathsf{Stash},  \mathsf{\textbf{T}},\textsf{st})$:} 

\noindent\textbf{Input:} the size of write batch $|\mathsf{W}|$ (public parameter), the bin buffer $\mathsf{Buf}$, the message stash $\mathsf{Stash}$, the set of oblivious message tables $\mathsf{\textbf{T}}$.
\begin{algorithmic}[1]
\State Choose $|\mathsf{W}|$ random distinct request $\mathsf{W} = \{k_i, v_i\}_{i = 1}^{|\mathsf{W}|}$
\If {$st.tempBin < k-1$}
\State $\mathsf{OBin} := \texttt{OHT.SimBuild}(\mathsf{W})$ \Comment{Create an oblivious bin with simulation}
\State $st.tempBin++$
\State Add $\mathsf{W}$ to $\mathsf{TempStash}$
\Else 
\State Remove $\{\mathsf{OBin}_i\}_{i = 1 ... k-1}$ from $\mathsf{Buf}$
\State $\mathsf{OBin} := \texttt{OHT.SimBuild}(\mathsf{W}\cup\mathsf{tempStash})$ \Comment{Create a larger $\mathsf{OBin}$}
\State $st.tempBin = 0$, and clear $\mathsf{TempStash}$
\EndIf

\State $idx\gets\mathsf{Buf}.\texttt{push}(\mathsf{OBin})$
\State Add $(\mathsf{W}, idx)$ to $\mathsf{Stash}$
\If{$|\mathsf{Stash}| \geq m$}   
\State \Call{OblMerge}{$\mathsf{Stash}$} \Comment{Proceed on the background}
\EndIf
\Procedure{OblMerge}{$\mathsf{Stash}$}
\State Extract $\mathsf{Stash}$ to  $\{\mathsf{W}_i\}_{i = 1...m}, I = \{idx_i\}_{i = 1...m}$
\State $\mathsf{OMT} := \texttt{OHT.SimBuild}(\bigcup^{m}_{i=1}W_i)$
\State Add $\mathsf{OMT}$ to $\mathsf{\textbf{T}}$
\State Remove $\{\mathsf{OBin}_i\}_{i \in I}$ from $\mathsf{Buf}$
\EndProcedure
\State Update $\textsf{st}$
\end{algorithmic}

\vspace{2pt}
\underline{$\mathsf{R'} \leftarrow \texttt{SimPong.OblRead}( |\mathsf{R}|, \mathsf{Buf},  \mathsf{\textbf{T}},  \textsf{st})$:} 

\noindent\textbf{Input:} the public parameter $|\mathsf{R}|$, the bin buffer $\mathsf{Buf}$, and the set of oblivious message tables $\mathsf{\textbf{T}}$.
\begin{algorithmic}[1]
\State Create $|\mathsf{R}|$ random read requests in the form $\{k_i, v_i\}$
\For {$(k_i, v_i) \in \mathsf{R}$}
\State Initialize a global boolean tag $\textsf{found} = \textsf{false}$. 
\For{$\mathsf{OBin} \in \mathsf{Buf}$ } 
\State $k = \texttt{SimOblChoose}(\textsf{found}, k, \perp)$
\State $v' = \mathsf{OBin.}\texttt{SimLookup}(k)$
\State $\textsf{OHTFound} = v'!= \perp $
\State $v = \texttt{SimOblChoose}(\textsf{OHTFound}, v', v)$
\State $\textsf{found} = \textsf{OHTFound} \vee \textsf{found}$
\EndFor  
\For{$\mathsf{OMT} \in \mathsf{\textbf{T}}$ } 
\State $k = \texttt{SimOblChoose}(\textsf{found}, \perp, k)$
\State $v' = \mathsf{OMT.}\texttt{SimLookup}(k)$
\State $\textsf{OHTFound} = v'!= \perp $
\State $v = \texttt{SimOblChoose}(\textsf{OHTFound}, v', v)$
\State $\textsf{found} = \textsf{OHTFound} \vee \textsf{found}$
\EndFor  
\State  Append $(k, v)$ to $\mathsf{R'}$. 
\EndFor
\end{algorithmic}

\end{multicols}
\end{tcolorbox}

  \caption{{\nameM{}}'s and its simulator algorithms for the oblivious message store.}\label{fig:SimHom}    
\end{figure*}

Next, we prove the communication trace indistinguishability of the metadata-private message store \nameM{}. Note that \nameM{} adopts standard oblivious algorithms (e.g., $\texttt{OHT.Build}$, $\texttt{OHT.Lookup}$ and $\texttt{OblChoose}$), we assume the existence of simulators $\texttt{OHT.SimBuild}$, $\texttt{OHT.SimLookup}$ and $\texttt{SimOblChoose}$ for ease of presentation.

\begin{myTheorem}\label{them:pong}
Given an oblivious hash table $\mathsf{OHT}$, an oblivious choose algorithm $\mathsf{OblChoose}$, an oblivious sort algorithm $\mathsf{OSort}$, and a secure encryption scheme $(\mathsf{Encrypt}$, $\mathsf{Decrypt})$, the metadata-private message store \nameM{} achieves communication trace indistinguishability as defined in Definition~\ref{def:ctu} in every communication round.
\end{myTheorem}

\noindent\textbf{Proof~}
From a high level, $\mathsf{Pong}$ empowers the client to send (write) and retrieve (read) messages through an oblivious message storage system. Similarly, we formulate the simulator algorithms for \nameM{}'s oblivious message store, as illustrated in Figure~\ref{fig:SimHom}, and demonstrate that the traces observed by the attacker in both the original and simulated algorithms are indistinguishable.  Below, we detail the indistinguishability of the read and write operations.

\noindent{\bf Write.}

\begin{itemize}
\item (Line 1) The original algorithm does not engage in processing, whereas the simulator randomly generates a batch of write requests of the same size as the original one. Given that they share the same size for write requests, they will be created to an oblivious bin of the same size.
\item (Lines 3, 8, and 18) By the security of oblivious hash tables, the algorithms \texttt{OHT.Build} and the corresponding simulator algorithm \texttt{OHT.SimBuild} process the indistinguishable memory access patterns.
\item (Lines 4-7, 9, and 11-12) These lines execute the same operations, making it impossible for the attacker to distinguish between the two experiments based on them.
\end{itemize}

\noindent{\bf Read.}

\begin{itemize}
\item (Line 1) 
The original algorithm does not perform any actions, whereas the simulator algorithm creates a set of read requests with the same size as the original one.
\item (Line 2) Both algorithms scan all the query requests to identify the reading task. Since the read requests have the same size, it is impossible to distinguish between the original algorithm and the simulator algorithm.
\item (Lines 4-10) These lines traverse through all $\mathsf{OBin}$s to find the matching results. During this procedure, due to the adoption of $(\mathsf{OHT.Lookup}$, $\mathsf{OHT.SimLookup})$ and \\ $(\mathsf{OblChoose}, \mathsf{SimOblChoose})$, they generate indistinguishable memory access patterns.
\item (Lines 11-17) These lines repeat lines 4-10 except that it looks up the oblivious message tables. Similarly, due to the oblivious lookup function of the oblivious hash table, they generate indistinguishable memory access patterns.
\item (Line 18) This step is identical in both algorithms.
\end{itemize}
Since all write and read requests are processed with indistinguishable memory access patterns, the attacker cannot differentiate between \nameM and the simulator algorithms. In other words, the attacker cannot determine where a write request is located in the storage and which clients requested it later, making it unable to distinguish different traces based on access patterns.

According to Theorem~\ref{them:ping} and Theorem~\ref{them:pong}, both \nameN and \nameM achieve communication trace indistinguishability within a single round. 
Namely, for any client connected to each system, the attacker cannot identify which two clients are communicating. Furthermore, the attacker cannot determine whether a client is online as idle or actively engaged in a real communication conversation. Since the observable interactions with \nameN and \nameM are all content-independent, \name achieves  communication trace indistinguishability in one round. 

Next, we continue to prove that the indistinguishability still works for multi-round cases. In scenarios where all clients remain online, the result is straightforward. However, our primary focus is on more general cases where clients are online/offline at a variable rate. Without loss of generality, we assume that all clients are online at fixed rate, and the status of being idle or active is unknown to the attacker.

\begin{myTheorem}
Assume that the scheme $(\mathsf{Encrypt}, \mathsf{Decrypt})$ is CPA secure and key-private, $\mathsf{PRF}$ is secure pseudo-random function, $\mathsf{OHT}$ is a two-layer oblivious hash table, and the adopted oblivious building blocks $\mathsf{OSort}, \mathsf{OblChoose}$ are
correct and private. Then, the system \name{} achieves communication trace indistinguishability except a negligible probability.
\end{myTheorem}
\noindent\textbf{Proof~} To establish communication trace indistinguishability, the key is to demonstrate that the communication pattern and the memory access pattern are indistinguishable. The oblivious notification system \nameN and the oblivious storage system \nameM{} effortlessly ensure indistinguishable memory access patterns. Thus, the focus of our argumentation lies in asserting that the communication patterns of all clients remain indistinguishable across multiple rounds.

Note that whether clients are engaged in communication is independent of their online status. As outlined in \S\ref{sec:schedule}, \name{} introduces idle clients to send and receive messages and caches them at the proxy if the receiver is offline. Additionally, each client will receive a message even if they are not the intended receiver. In other words, in each round, any client in the set has the potential to be his/her buddy. Consequently, in multiple rounds, all clients connected to the system have an equal possibility of being the receiver. This completes the proof.

\subsection{Case Study: Evaluation on CollegeMsg Metadata Dataset} \label{sec:evaluation_framework}

To assess the efficiency of the notify-before-retrieval framework compared to the traditional dial-before-conversation framework, we show a case study using the CollegeMsg temporal network dataset from Stanford SNAP\footnote{\url{https://snap.stanford.edu/data/CollegeMsg.html}}. This analysis aims to answer the question: why is the notify-before-retrieval framework more efficient than the dial-before-conversation framework in real-world messaging scenarios? 

\noindent\textbf{Dataset.} The CollegeMsg dataset contains anonymized chat metadata, consisting of tuples (sender, receiver, timestamp) from 1,899 users over a 193-day period, totaling 59,835 messages. This dataset provides a realistic sample of messaging patterns, including bursts of activity and frequent conversation switching, which are common in instant messaging contexts.

\noindent\textbf{Modeling the dial-before-conversation framework.}
The dial-before-conversation framework requires users to dial and establish a conversation session before messaging, restricting each user to one conversation at a time. If the recipient is busy, the sender must wait until the recipient becomes available. In instant messaging contexts, users frequently switch between conversations, potentially causing delays.

To simulate this framework, we divide the timestamps into fixed time windows (ranging from 30 seconds to 300 seconds in this evaluation). Each conversation occupies one such window, during which the recipient is unavailable for other conversations. For each recipient, we model a queue where only one conversation (i.e., the same sender) can be processed per window. If multiple senders target the same recipient within overlapping windows, subsequent messages are delayed to the next available window. To focus on waiting time rather than processing time, we set the dialing latency to 0 second and the conversation latency (message delivery time) to 0.5 seconds.

\noindent\textbf{Modeling the notify-before-retrieval framework.}
The notify-before-retrieval framework, as implemented in \name{}, eliminates the need for prior coordination. Senders notify recipients and store messages, allowing asynchronous retrieval without waiting for recipient availability. Based on our system evaluation, we assign an end-to-end processing latency of 3 seconds per message (as from results in Fig.~\ref{fig:ppe2eLatency}).

\begin{figure}
    \centering
    \includegraphics[width=0.95\linewidth]{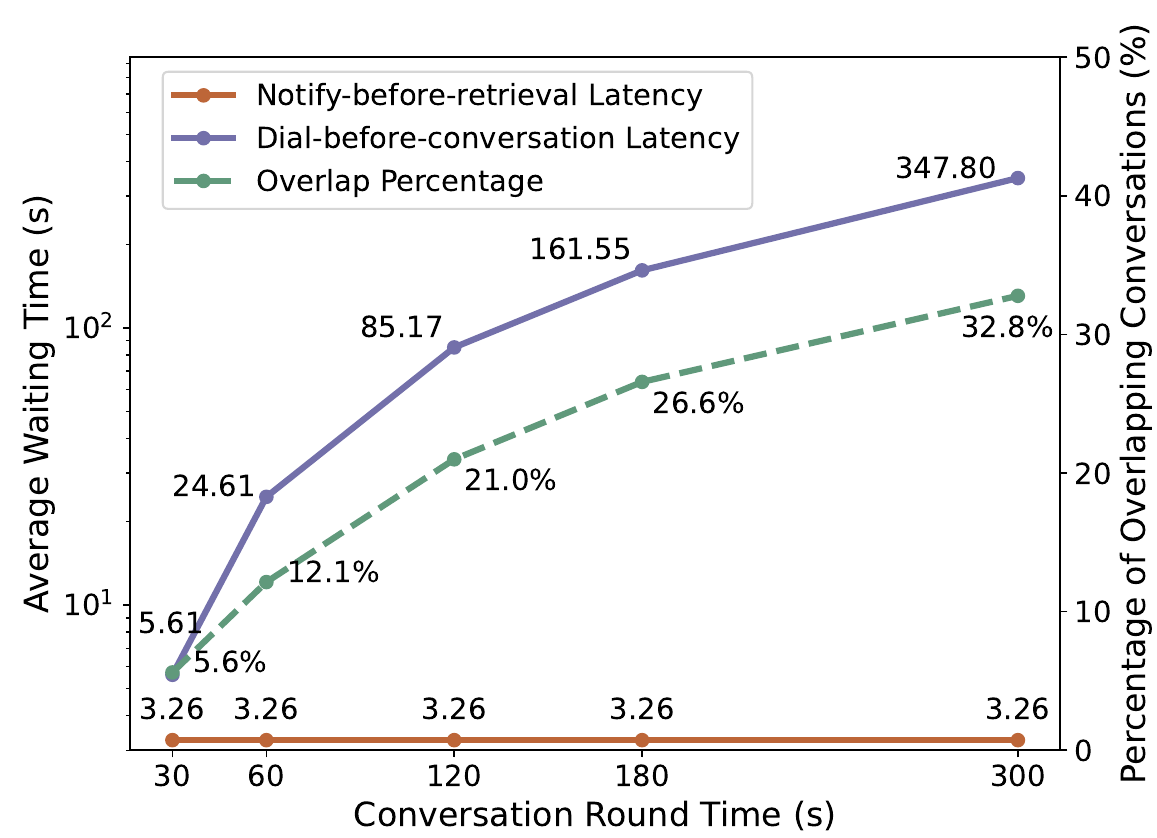}
    \caption{Evaluation on the two frameworks on CollegeMsg dataset.}
    \label{fig:eval_framework}
\end{figure}

\noindent\textbf{Results and analysis.} The results of our evaluation are presented in Fig.~\ref{fig:eval_framework}. As the conversation window size increases, the number of messages experiencing conflicts (i.e., overlapping conversation windows) also increases. With a window size of 300 seconds—a common parameter in prior work~\cite{osdi21Addra,sosp19yodel}—32.8\% conversations encounter waiting times due to overlapping conversation windows.

In the dial-before-conversation framework, the average latency across all messages is 347 seconds, with waiting time being the dominant factor due to recipients being unavailable during ongoing conversation sessions. In contrast, the notify-before-retrieval framework achieves an average latency of just 3.26 seconds, primarily reflecting processing time, with negligible waiting time.

These results highlight the efficiency of the notify-before-retrieval framework for instant messaging. In real-world scenarios, as captured by the CollegeMsg dataset—where users frequently switch conversations and send message bursts—the notify-before-retrieval framework reduces latency by orders of magnitude, offering a significantly more satisfactory messaging experience.

\end{document}